\documentclass[smallextended]{svjour3preprint}
\usepackage[dvips]{epsfig}

\usepackage{graphicx,latexsym, amsmath,amsfonts}
\usepackage{epsfig}

\begin{document}

\title{\bf{Comparing disease control policies for interacting wild populations}}
\author{Iulia~Martina~Bulai \and Roberto~Cavoretto \and Bruna~Chialva \and
Davide~Duma \and Ezio~Venturino}
\institute{I.M. Bulai \at
              Department of Mathematics \lq\lq Giuseppe Peano\rq\rq, University of Torino, via Carlo Alberto 10, I--10123 Torino, Italy \\
              \email{iuliam@live.it}           
            \and
         R. Cavoretto \at
              Department of Mathematics \lq\lq Giuseppe Peano\rq\rq, University of Torino, via Carlo Alberto 10, I--10123 Torino, Italy \\
              \email{roberto.cavoretto@unito.it}           
         \and
         B. Chialva      \at
              Department of Mathematics \lq\lq Giuseppe Peano\rq\rq, University of Torino, via Carlo Alberto 10, I--10123 Torino, Italy \\
              \email{bruna.chiara@alice.it}           
            \and
         D. Duma   \at
              Department of Mathematics \lq\lq Giuseppe Peano\rq\rq, University of Torino, via Carlo Alberto 10, I--10123 Torino, Italy \\
              \email{davideduma@gmail.com}           
            \and
         E. Venturino   \at
              Department of Mathematics \lq\lq Giuseppe Peano\rq\rq, University of Torino, via Carlo Alberto 10, I--10123 Torino, Italy \\
              \email{ezio.venturino@unito.it}           
}

\date{}

\maketitle

\abstract{
We consider interacting population systems of predator-prey type, presenting four models of control strategies for epidemics among the prey.
In particular to contain the transmissible disease, safety niches are considered, assuming they lessen the disease spread,
but do not protect prey from predators. This represents a novelty with respect to standard ecosystems
where the refuge prevents predators' attacks. The niche is assumed either to protect the healthy individuals,
or to hinder the infected ones to get in contact with the susceptibles, or finally to reduce altogether
contacts that might lead to new cases of the infection. In addition a standard culling procedure is also analysed.
The effectiveness of the different strategies are compared.
Probably the environments providing a place where disease carriers
cannot come in contact with the healthy individuals, or where their contact rates are lowered,
seem to preferable for disease containment.
}

\section{Introduction}

In population models predator-prey and competition systems play a dominant role, since the blossoming of this discipline about a century ago.
In more recent times, more refined models try to better describe reality. Since prey try to seek protection against attacks of their predators
in the features of the environment, scientists have tried to incorporate this behavior into the interaction models.
Early contributions in this respect can be found in \cite{MS,K,M76}.
The introduction of refuges
has lead to the observation that the Lotka-Volterra models gets stabilized \cite{Gonzalez2012} even to show global asymptotic stability,
\cite{Collings1995,G03}. This shows the relevant role that spatial refuges exert in shaping
the dynamics of predator-prey interplay. The refuge is expressed in the equations by reducing the amount of prey population available for hunting
by the predators.

In this classical setting,
if $Y$ denotes the prey population that can take cover, by $Y_n$ we denote the number of individuals who find protection in the niches that are available for
their safety.
Thus there are only $Y-Y_n$ individuals that can interact with the predators. There could be several functional forms that can be chosen for $Y_n$.
The simplest one is a constant value, $Y_n=Y_0$, with $Y_0\in \textbf{R}_+$, or alternatively one could take a linear function of the prey population, $Y_n=Y_0 Y$,
\cite{Gonzalez2012} or also a linear function of
the predators $X$, $Y_n=Y_0 X$ \cite{Ruxton1995short}.


Ecoepidemiology investigates the influence of diseases in ecosystems, see Chapter 7 of \cite{MPV}. It appears therefore that the refuges for some
of the populations involved can be introduced also in this context. However, instead of using the environmental niches as protection against
the predators, i.e. as an ecological tool as described above, we employ them in order to investigate whether they can influence the disease spread,
i.e. we give them an epidemiological meaning. Therefore, it is not against predators that prey are protected, but we rather consider the case in
which the healthy prey for some reason due to the conformation of the environment can avoid
to come in contact with disease-carriers of their own population and therefore be somewhat protected from the epidemics. This is achieved
by reduced contact rates that they have with infected individuals.
Of all the various possible types of niche, to keep things simple, we just take the constant case, $Y_n=Y_0$.

In the next Sections, we present three models for the refuges and one for another common disease-control method,
namely culling,
based on the ecoepidemic system presented in \cite{EV95}. The first three
differ in the way the refuge is modeled. In Section 2, some of the susceptibles are prevented
from interaction with infected individuals. In Section 3, it is part of the infected that are unable to become in contact with healthy individuals.
In Section 4, we look at a reduced contact rate. Section 5 contains the analysis of the culling strategy.
After a brief discussion of bistability of some equilibria, the final Section compares the findings.

\section{The 
refuge for the healthy prey}

Consider at first the system in which the susceptibles are stronger and therefore
able to reach places unattainable by the
diseased individuals, because these
indeed are weakened by the disease. Thus the infectious individuals
cannot come in contact with the healthy remote individuals, and therefore cannot infect them.
Let $s$ denote
the fixed number of susceptibles that escape from the spread of the epidemics using the refuge.

The model is formulated as follows. The healthy prey $R$ reproduce with net reproduction rate $a$, are subject to intraspecific competition
only with other sound individuals at rate $b$
and are hunted by predators at rate $c$. Those that can be infected by the diseased prey individuals $U$, as discussed above,
leave their class at rate $\lambda$, to enter into the class of sick inviduals. The latter do not reproduce, are hunted at a rate $k\ne c$
by the predators. Here $k>c$ means that they are weaker than sound ones, and therefore easier to capture, while $k<c$ instead takes into
account the fact that they might be less palatable than the healthy ones. Finally, they can recover the disease at rate $\omega$ and therefore
reenter into the $R$ population. As mentioned above, infected are assumed not to contribute to intraspecific pressure, either of sound prey
or among themselves; this again is grounded in the fact that their disease-related weakness prevents them to compete with the other individuals
in the population.
The predators are assumed to have also other food sources, for which they reproduce at rate $d$, but clearly
get a benefit from the interactions with the healthy prey expressed by the parameter $e<c$. This constraint expresses the fact that
the amount of food they get from the captured prey cannot exceed its mass. So far all
the system parameters are nonnegative.
For the predators hunting the infected prey, instead,
we could model two different situations. For $h>0$, the infected cause a damage to the predators, killing them.
In this paper we concentrate only on this case. In the opposite case we could have the
normal situation in which predators get a reward from capturing the diseased prey, so that in this situation we would have $0<-h<k$.
In summary, the ecoepidemic model with inclusion of a disease-safety niche for the susceptibles reads
\begin{eqnarray}\label{mod_2_susceptibles}
\frac{dR}{dt}&=&R[a-bR-cF] + \omega U -\lambda \max\{0,(R-s)\}U,\\ \nonumber
\frac{dU}{dt}&=&\lambda \max\{0,(R-s)\}U - U[ kF + \omega]- \mu U , \\ \nonumber
\frac{dF}{dt}&=&F[d+eR-fF-hU] .
\end{eqnarray}

When $R<s$, the last term in
the first equation and the first one in the second equation vanish,
the maximum function preventing them to
provide positive and negative contributions to these equations respectively, which makes no sense biologically.
It follows also that for $R<s$ the infected prey in the system disappear,
since in the second equation the term on the right hand side is always negative.
Thus the system settles to one of the equilibria of the classical
disease-free predator-prey model, with logistic correction for the prey alternative food supply for the predators, see \cite{EV95} for its brief analysis.
For the benefit of the reader a short summary of its findings is presented also here at the top of Section \ref{comparison}.

\subsection{Equilibria}
The equilibria $P_k=\left(R_k,U_k,R_k\right)$ of (\ref{mod_2_susceptibles}) are $P_1=\left(0,0,0\right)$,
$P_2=\left(0, 0,d f^{-1} \right)$,
$P_3=\left(a b^{-1}, 0, 0 \right)$,
\begin{eqnarray*}
P_4=\left({af-cd \over bf+ce}, 0, {ae+bd \over bf+ce} \right),\quad
P_5&=&\left({\lambda s+ \omega + \mu \over \lambda}, {(a-bR_5)R_5 \over \mu },0  \right).
\end{eqnarray*}
The first three points are always feasible, $P_4$ is feasible for
\begin{equation}\label{P4_feas}
af\ge cd,
\end{equation}
and $P_5$ is whenever $a\ge bR_5$, i.e. for
\begin{equation}\label{P5_feas}
b(\lambda s+ \omega + \mu)\le a\lambda.
\end{equation}
Then there is coexistence $P_6=\left(R_6, U_6, F_6\right)$.
Its population values are obtained solving for $F$ and $U$ respectively the second and third equations in (\ref{mod_2_susceptibles}), taking obviously $R_6>s$,
thus giving
$$
F_6=\frac 1k \left[ \lambda (R_6-s) - \omega -\mu\right], \quad U_6=\frac 1h \left[ d+eR_6 - f F_6 \right].
$$
Note that this makes sense only for $R_6>s$, since otherwise the second equilibrium equation gives either
$U_6=0$ or $F_6=-\omega k^{-1}$, but both results are in contrast with coexistence.
Substituting into the first one, we obtain the quadratic equation $W(R)\equiv \sum _{k=0}^2 a_k R^k=0$ whose roots give the values of $R_6$.
Its coefficients have the following values
\begin{eqnarray*}
a_2 = \frac {\lambda}h \left( \frac fk \lambda - e\right) -b -\frac ck \lambda, \quad a_0=\frac 1{hk} \left( dk+fs\lambda+f(\omega +\mu) \right) (s\lambda+\omega),\\
a_1=a+\frac ck (s\lambda+\omega+\mu) +\frac 1{hk} [ (s\lambda+\omega) (ek-f\lambda) - \lambda (dk+fs\lambda +f(\omega+ \mu)) ].
\end{eqnarray*}
Now, since $a_0>0$, if the parabola $W(R)$ is concave one positive root will exist. Thus a sufficient condition for the existence of $P_6$ is
$a_2<0$, i.e., explicitly,
\begin{eqnarray}\label{hatP5_exist}
f\lambda ^2 < e\lambda k+h[bk +c\lambda].
\end{eqnarray} 
For feasibility, we need also the other population values at a nonnegative level, a fact which is attained for $U_6$ if $ek>f\lambda$, else
we must impose it 
\begin{eqnarray}\label{hatP5_feas1}
s < R_6 < \frac {dk+fs\lambda +f(\omega+\mu)} {f\lambda -ek}
\end{eqnarray} 
as we do for $F_6$ to obtain
\begin{eqnarray}\label{hatP5_feas2}
R_6 > s + \frac {\omega+\mu}{\lambda}.
\end{eqnarray} 

\subsection{Stability}
Denoting as usual by $H(x)$ the Heaviside function, $H(x)=1$ for $x>0$, $H(x)=0$ for $x\le 0$,
the Jacobian of (\ref{mod_2_susceptibles}) is
$$
J=
\left[
\begin{array}{ccc}
a-2bR-\lambda H(R-s) U-cF & -\lambda \max\{0,(R-s)\}+\omega & -cR \\
\lambda H(R-s) U & \lambda \max\{0,(R-s)\}-kF-\omega- \mu & -kU \\
eF & -hF& J_{33}
\end{array}
\right],
$$
$J_{33}= d+eR-hU-2fF $.
The eigenvalues for $P_1$ are
$-\omega- \mu$, $d$, $a$, entailing its instability.
Those for $P_2$ are
$-(dk+f(\omega+ \mu))f^{-1}$, $-d$, $(af-cd)f^{-1}$
giving the stability condition
\begin{eqnarray}\label{P2_stab}
af<cd.
\end{eqnarray}
Comparing this condition with (\ref{P4_feas}), we observe that there is a transcritical bifurcation, for which $P_4$ emanates from $P_2$ when the
latter becomes unstable. In other words, introducing the healthy prey invasion number
\begin{eqnarray}\label{HPIN}
R^{(i)}\equiv \frac {af}{cd},
\end{eqnarray}
we have that for $R^{(i)}>1$ the healthy prey establish themselves in the environment.

For $P_3$ the eigenvalues are
$(bd+ae)b^{-1}$, $(\lambda a-\lambda s b-b(\omega+ \mu))b^{-1}$, $-a$,
giving instability.

\begin{figure}[!hb]
\centering
\includegraphics[width=10cm]{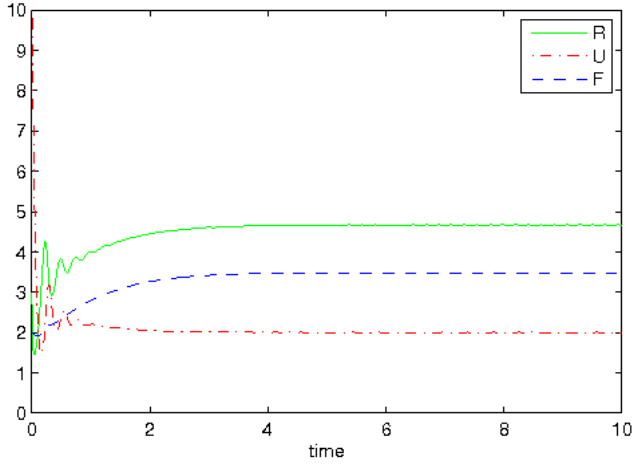}
\caption{The coexistence equilibrium is stably attained for the following choice of parameters:
$a=21$, $b=0.3$, $c=1$, $d=1$, $e=0.5$, $f=0.9$, $h=0.1$, $k=10$, $\lambda=10.2$, $\omega=0.8$, $\mu=2.8$, $s=0.9$.}
\label{fig-P5}
\end{figure} 

At $P_4$ 
one eigenvalue is easily factored out, 
$$
\frac{\lambda (af-cd)-k(bd+ae)}{ce+bf}-\lambda s-\omega-\mu,
$$
while the remaining ones are roots of the quadratic equation
\begin{equation}\label{q1}
T(\delta)=\delta^2+b_1\delta+b_2=0,
\end{equation}
where letting $D=ce+bf$,
\begin{eqnarray*}
b_1 & = & \frac{t_1}{D},\quad b_2=\frac{t_3}{D},\quad t_1=af(b+e)+bd(f-c)\\
t_3 & = & (bd+ae)(af-cd),\quad t_2=t_1^2+4t_3(bf+ce).
\end{eqnarray*}
Explicitly,
\begin{eqnarray}\label{quadr}
T_{1,2}=\frac{-b_1\pm\sqrt{b_1^2-4b_2}}{2}=\frac{t_1\pm\sqrt{t_2}}{2(ec+bf)}.
\end{eqnarray}
By the feasibility condition (\ref{P4_feas}),
$t_3<0$ so that $t_2<t_1^2$. Hence both roots of (\ref{quadr}) have negative real part. Stability hinges then just on the first
eigenvalue, i.e. $\lambda R_4<kF_4+\lambda s+\omega+\mu$ or explicitly the following condition
\begin{eqnarray}\label{P4_stab}
\lambda\frac{af-cd}{bf+ce}<k\frac{ae+bd}{bf+ce}+\lambda s+\omega+\mu.
\end{eqnarray}

An eigenvalue of $ P_5$ is $d+e  R_5-h  U_5$, the remaining ones
are the roots of $T(\theta) = \theta^2-c_1 \theta+c_2=0$, with
\begin{eqnarray*}
c_1 = a-2b R_5-\lambda U_5, \quad c_2 = \mu \lambda  U_5>0.
\end{eqnarray*}
Explicitly,
$$
T_{1,2}=\frac{c_1\pm\sqrt{c_1^2-4c_2}}{2}.
$$
By Descartes' rule of signs, both have negative real part if $a<2b R_5+\lambda  U_5$.
But this inequality always holds, since 
\begin{eqnarray*}
&&
\frac{a-2b  R_5}{\lambda}
=\frac{a\lambda- 2b (\lambda s + \omega +\mu)}{\lambda^2} =\\
&&=\frac{\mu U_5}{\lambda s + \omega +\mu}- \frac{b(\lambda s + \omega +\mu)}{\lambda^2}
< U_5-\frac{b(\lambda s + \omega +\mu)}{\lambda^2}<U_5.
\end{eqnarray*}
Stability then hinges only on the first eigenvalue
\begin{equation}\label{P5_stab}
 U_5> \frac{d+e  R_5 }{h}.
\end{equation}

For the coexistence equilibrium $P_6$, we have run some simulations to show
not only that it satisfies the feasibility conditions (\ref{hatP5_feas1}) and
(\ref{hatP5_feas2}), but that it can be attained at a stable level.
Figure \ref{fig-P5} shows 
one such instance, for the hypothetical parameter values $s=0.9$ and
\begin{eqnarray}\label{param_val}
a=21, \quad b=0.3, \quad c=1, \quad d=1, \quad e=0.5, \quad f=0.9, \\ \nonumber
h=0.1, \quad k=10, \quad \lambda=10.2, \quad \mu=2.8, \quad \omega=0.8. 
\end{eqnarray}
Here the $R_6$ equilibrium value is
much higher than the number of individuals $s$ that can take cover in the safety niche.
Observe also that the same inequality holds also for all the healthy prey population values before attaining the equilibrium level.

\section{The cover for the infected}

Assume now that part of the infected are somehow confined in an environment in which healthy prey cannot enter. In this way the contagion
risk is reduced. Let $p$ denote the fixed number of infected that inhabit the unreacheable territory. With the remaining notation similar
to model (\ref{mod_2_susceptibles}), the system in our present case reads
\begin{eqnarray}\label{mod_2_infected}
\frac{dR}{dt}&=&R[a-bR-cF-\lambda \max\{0,(U-p)\}] + \omega U ,\\ \nonumber
\frac{dU}{dt}&=&\lambda \max\{0,(U-p)\}R - U[ kF + \omega]- \mu U , \\ \nonumber 
\frac{dF}{dt}&=&F[d+eR-fF-hU] .
\end{eqnarray}

Again, here we have to remark that for
$U<p$ the contributions to the infected class is to be understood to drop to zero. In such case, once again,
the infected prey in the system vanish,
and the system settles to any equilibrium of the classical
disease-free predator-prey model, \cite{EV95}.

\subsection{Equilibria}
For (\ref{mod_2_infected}) the equilibria are again the origin $\widetilde P_1\equiv P_1 =\left(0,0,0\right)$ and the point $\widetilde P_2\equiv P_2$,
while the healthy prey thrives again at $\widetilde P_3\equiv P_3$,
coexistence of healthy prey and predators is attained at level
$\widetilde P_4\equiv P_4$ and the predator-free point
\begin{eqnarray*}
\widetilde P_5=\left( \widetilde R_5,{1\over \mu} \widetilde R_5 (a-b\widetilde R_5 ),0  \right),
\end{eqnarray*}
where $\widetilde R_5$ solves the quadratic equation
$$
b\lambda R^2 -R [a\lambda +b(\omega+\mu)] +a (\mu+\omega) -p\lambda=0.
$$
In view of the convexity of this parabola, there is exactly one positive root if
\begin{equation}\label{tR5}
a (\mu+\omega) <p\lambda,
\end{equation}
while there are two such positive roots if
\begin{equation}\label{tR5R5}
a (\mu+\omega) >p\lambda, \quad 
[a\lambda +b(\omega+\mu)]^2 > 4b\lambda [a (\mu+\omega) -p\lambda].
\end{equation}
In addition $\widetilde P_5$ is feasible for the condition
\begin{eqnarray}\label{tP5_feas}
\widetilde R_5 \le \frac a b.
\end{eqnarray}

The presence of the coexistence equilibrium $\widetilde P_6=(\widetilde R_6,\widetilde U_6,\widetilde F_6)$ can be discussed as follows. We take $U>p$, else the second equilibrium
equation of (\ref{mod_2_infected}) cannot be solved for
positive values of the populations.
From the last equilibrium equation of (\ref{mod_2_infected}) we solve for $F$ obtaining
$$
\widetilde F_6=\frac 1f (d+eR-hU)
$$
and substitute into the remaining equations to obtain two conic sections
\begin{eqnarray*}
\Psi(R,U)\equiv -\left( b+\frac cf e\right) R^2 + \left( \frac cf h - \lambda \right) RU + \left( p\lambda -\frac cf d+a\right) R +\omega U=0,\\
\Phi(R,U)\equiv \frac kf hU^2+ \left(\lambda - e\frac kf\right) RU - \left(\frac kfd +\omega +\mu \right)U -p \lambda R=0,
\end{eqnarray*}
of which we seek an intersection $(\widetilde R_6,\widetilde U_6)$ in the first quadrant.
We study each one of them separately.

The implicit function $\Phi=0$ can be solved as a function $R = \rho(U)$,
$$
\rho(U) \equiv U \frac {kh U - \left[ f(\omega +\mu)+dk \right]}{fp\lambda +(ek-f\lambda)U} .
$$
The function has a zero at the origin and another one at $U^{0}=[f(\mu+\omega)+kd] (hk)^{-1}>0$. It
has also a vertical asymptote at $U^{\infty}=fp\lambda (f\lambda-ek)^{-1}$.
Asymptotically, for large $U$, we find
\begin{equation}\label{slope_rho}
\rho(U) \sim \alpha U \equiv \frac {hk}{ek-f\lambda} U.
\end{equation}
We can rewrite $\rho$ as follows, and then compute its second derivative:
$$
\rho(U) = \alpha \frac {U-U^0}{U^{\infty}-U^0} U, \quad 
\rho ''(U) = -2\alpha U^{\infty}\frac {U^0-U^{\infty}} {(U-U^{\infty})^3}.
$$
Observe that $\alpha >0$ if and only if $U^{\infty}<0$.
There are three possible situations that can arise, depending on the sign of $U^{\infty}$.
\begin{itemize}
\item [(A)] $U^{\infty}<0<U^0$; in this case there is a feasible branch mapping $[U^0,+\infty )$ surjectively onto
$[0, \infty )$; the feasible branch of $\rho(U)$ is increasing; the function is convex for $U>U^{\infty}$ and
thus the whole feasible branch is.
\item [(B)] $0<U^{\infty}<U^0$; in this case there is a feasible branch mapping $(U^{\infty},U^0 ]$ surjectively onto
$[0, +\infty )$; the feasible branch of $\rho(U)$ is decreasing; the function is convex for $U>U^{\infty}$ and
thus the whole feasible branch is.
\item [(C)] $0<U^0<U^{\infty}$; in this case there is a feasible branch mapping $[U^0,U^{\infty })$ surjectively onto
$[0, +\infty )$; the feasible branch of $\rho(U)$ is increasing; the function is convex for $U<U^{\infty}$ and
thus the whole feasible branch is.
\end{itemize}
The inverse function $U=\rho^{-1}(R)$ maps $[0,+\infty)$ surjectively onto $[U^0,+\infty )$, $(U^{\infty},U^0 ]$
and $[U^0,U^{\infty })$ respectively in each case (A), (B), (C).

We proceed similarly with the implicit function $\Psi(R,U)=0$, rewriting it as $U=\xi (R)$,
$$
\xi(R) \equiv R \frac {(bf+ce)R + cd-af-fp\lambda }{\omega f +(ch-f\lambda)R} .
$$
It has a zero at $R^0=(af+fp\lambda -cd)(bf+ce)^{-1}$,
a vertical asymptote at $R^{\infty}=\omega f (f\lambda-ch)^{-1}$ and asymptotically it behaves like a straight
line,
\begin{equation}\label{slope_xi}
\xi(R) \sim \gamma R \equiv \frac {bf+ce}{ch-f\lambda} R.
\end{equation}
Rewrite it again in compact form, so that
$$
\xi(R) = \gamma R \frac {R-R^0}{R-R^{\infty}}, \quad
\xi''(R) = -2\gamma R^{\infty} \frac {R^0-R^{\infty}}{(R-R^{\infty})^3}.
$$
Here $\gamma >0$ if and only if $R^{\infty}<0$.
In this case, more alternatives arise, since here also $R^0$ can be negative. We list them as follows:
\begin{itemize}
\item [(I)] $R^{\infty}<R^0<0$; there is an increasing feasible branch mapping $[0,+\infty )$ surjectively onto
$[0, \infty )$; the feasible branch is convex.
\item [(II)] $R^0<R^{\infty}<0$; as for (I) there is an increasing feasible branch mapping $(0,+\infty )$
surjectively onto $[0, +\infty )$; the feasible branch is concave.
\item [(III)] $R^0<0<R^{\infty}$; the feasible branch is increasing and maps $[0,R^{\infty })$ surjectively onto
$[0, +\infty )$; the feasible branch is convex.
\item [(IV)] $R^{\infty}<0<R^0$; the increasing feasible branch maps here $[R^0,+\infty )$ surjectively onto
$[0, \infty )$; the feasible branch is convex.
\item [(V)] $0<R^0<R^{\infty}$; in this case there is an increasing
feasible branch mapping $[R^0, R^{\infty})$ surjectively onto $[0, +\infty )$; the feasible branch is convex.
\item [(VI)] $0<R^{\infty}<R^0$; the is feasible branch decreases, mapping $(R^{\infty },R^0]$ surjectively onto
$[0, +\infty )$; the feasible branch is convex.
\end{itemize}

The coexistence equilibrium is represented by the intersections of $\rho^{-1}$ and $\xi$. Now in view of the
surjectivity and the continuity of these functions, whenever one vertical asymptote, either $U^{\infty}$ or
$R^{\infty}$ is feasible, the intersection is guaranteed. The only cases that are questionable are (A)-(I),
(A)-(II) and (A)-(IV). In these cases we compare the asymptotic behaviors of the two functions. To guarantee
an intersection, we need to have $\alpha^{-1} < \gamma$, comparing (\ref{slope_rho}) and (\ref{slope_xi}).
This condition becomes
\begin{equation}\label{slopes}
bhk+ek\lambda + ch\lambda > f \lambda ^2.
\end{equation}
Now case (A)-(I) and (A)-(II) both correspond to
$$
U^{\infty}<0, \quad R^0<0, \quad R^{\infty}<0,
$$
while (A)-(IV) gives the same situation with only the second above inequality reversed. Combining the two,
we are left with the first and the third conditions, namely 
$$
ek>f\lambda, \quad f\lambda<ch.
$$
Use of these into (\ref{slopes}) shows that the inequality is always satisfied,
$$
bhk+ek\lambda + ch\lambda - f \lambda ^2>bhk+ek\lambda >0.
$$
Hence a feasible intersection
exists also in these cases.

Uniqueness follows in view of the convexity properties of the feasible branches of the functions
$\rho^{-1}$ and $\xi$.

\vspace{0.3cm}
We have thus shown the following result.

\vspace{0.3cm}

{\textbf{Theorem}}. The feasible coexistence equilibrium $\widetilde P_6$ always exists and is unique.

\vspace{0.3cm}
\subsection{Stability}
The Jacobian of (\ref{mod_2_infected}) is
$$
\widetilde J = 
\left[
\begin{array}{ccc}
a-2bR-cF-\lambda \max\{0,(U-p)\} & -\lambda R H(U-p)+\omega & -cR \\
\lambda \max\{0,(U-p)\} & \lambda R H(U-p)-kF-\omega-\mu & -kU \\
eF & -hF & \widetilde J_{33}
\end{array}
\right] ,
$$
$\widetilde J_{33}=d+eR-hU-2fF$.

$\widetilde P_1$ is always unstable, since the eigenvalues are $a$, $d$ and $-\omega-\mu$.

For $\widetilde P_2$ we find the eigenvalues $-d$, $-(dk+f\omega+f\mu)f^{-1}$ and $(af-cd)f^{-1}$, giving
again the stability condition (\ref{P2_stab}).

\begin{figure}[!ht]
\centering
\includegraphics[width=5.8cm]{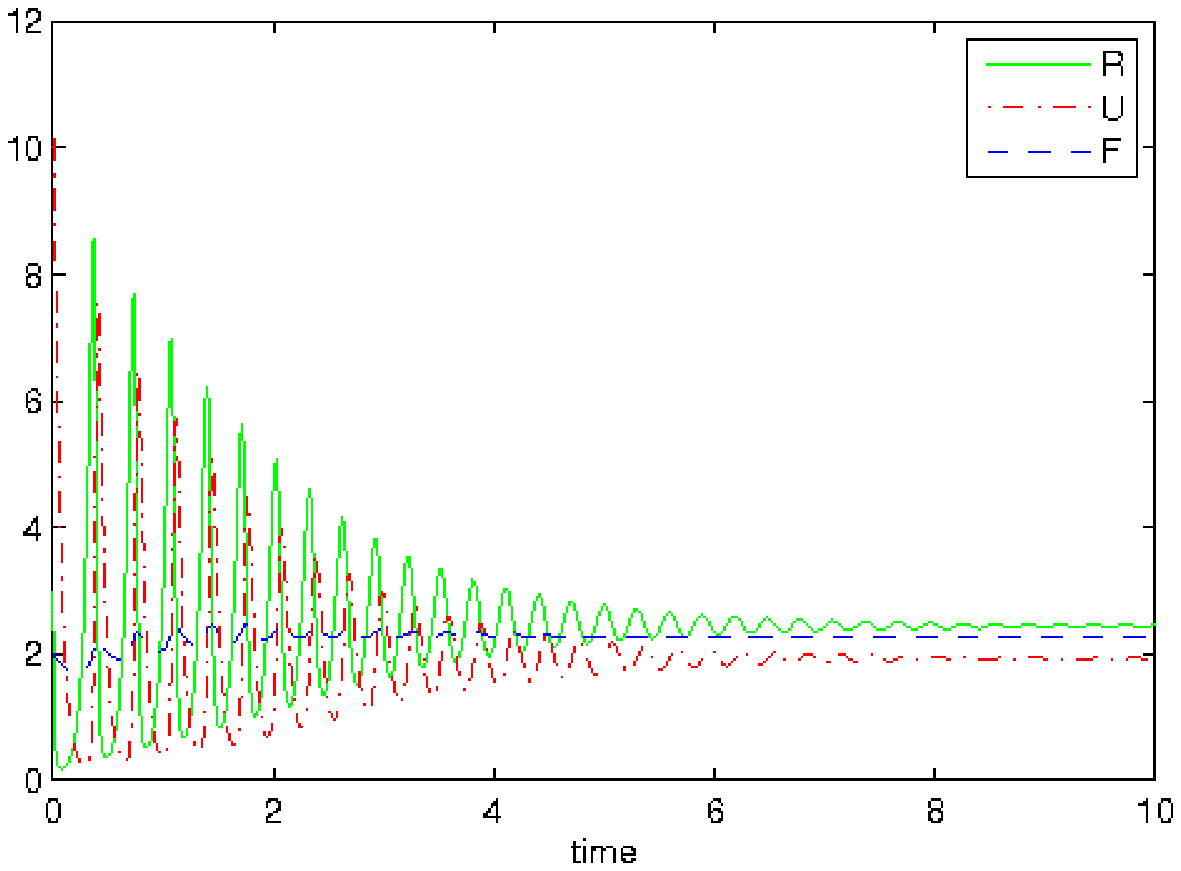}
\includegraphics[width=5.8cm]{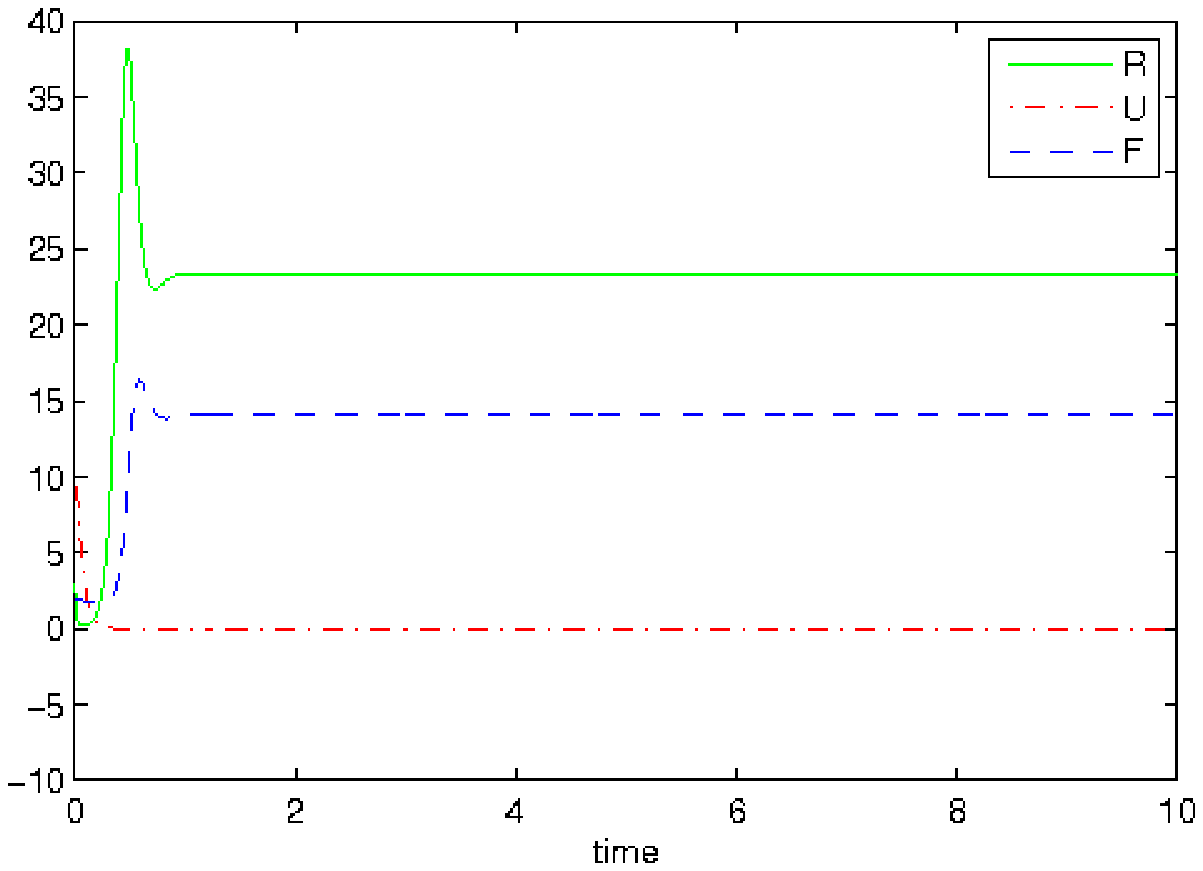}
\caption{Left: The coexistence equilibrium $\widetilde P_5$ is achieved when $\mu=0.28$ and $p=0.1$ and the remaining parameters are given by
(\ref{param_val}) as in Figure \ref{fig-P5}.
Right: The disease-free equilibrium is attained for $\mu=0.28$ and $p=0.4$ with the remaining parameters given by
(\ref{param_val}) as in Figure \ref{fig-P5}.
Note that the diseased population $U$ falls below the level $p$ very soon, and consequently both the healthy prey first and subsequently
the predators pick up, and finally settle to the coexistence equilibrium of the underlying demographic model.}
\label{tildeP5}
\end{figure} 

The point $\widetilde P_3$ is unstable, in view of the eigenvalues $-a$, $-\omega-\mu$,
$(ae+bd)b^{-1}>0$. 

For $\widetilde P_4$ we find the eigenvalue $-k \widetilde F -\omega-\mu<0$;
the Routh-Hurwitz conditions on the remaining minor of $\widetilde J$ are satisfied,
the determinant being $ce \widetilde R_4 \widetilde F_4>0$, the trace instead leading to $-(ae+bd)f(bf+ce)^{-1}<0$.
Thus, when feasible, $\widetilde P_4$ is unconditionally stable.

For the point $\widetilde P_5$ the Jacobian factorizes to give one explicit eigenvalue, from
which the first stability condition can be obtained,
\begin{equation}\label{tP5_stab1}
d+e \widetilde R_5 < h \widetilde U_5,
\end{equation}
and a quadratic equation, for which the Routh-Hurwitz criterion provides the remaining stability conditions
\begin{equation}\label{tP5_stab2}
b \widetilde R_5^2 +(\omega -\lambda \widetilde R_5) \widetilde R_5 + \omega \widetilde U_5>0, \quad
(b \widetilde R_5^2 +\omega \widetilde U_5) (\omega -\lambda \widetilde R_5) + \omega \lambda^2 \widetilde R_5^2
( \widetilde U_5 -p) >0.
\end{equation}

With the help of some simulations we can show that the coexistence equilibrium can be stably achieved, Figure \ref{tildeP5} left. The
refuge parameter used is $p=0.1$ while all the remaining ones are those (\ref{param_val}) as in Figure \ref{fig-P5}.
Note that in this case raising the niche level to 
$p=0.4$ causes the infected population at some point to fall below this threshold,
so that they are wiped out, Figure \ref{tildeP5} right. So while we stated that the disease-free point is not an equilibrium of (\ref{mod_2_infected}) per se,
in suitable situations it would certainly occur.
In fact when the infected population $U$ becomes smaller than the level $p$, and this occurs pretty early in the simulation
as observed in Figure \ref{tildeP5} right,
the sound prey first and then also
the predator populations suddenly surge to finally settle to the coexistence equilibrium of the underlying demographic model.

\section{The reduced contacts}

We consider now another situation, in which we assume that it is the rate of contacts between infected and susceptibles
that gets somewhat reduced, due to the effect of a protective niche.
In this case then we introduce the fraction $0\leq q \leq 1$ of avoided contacts. The model, using again the very same previous notation,
now becomes
\begin{eqnarray}\label{mod_2}
\frac{dR}{dt}&=&R[a-bR-cF-(1-q)\lambda U] + \omega U ,\\ \nonumber       
\frac{dU}{dt}&=&U[(1-q)\lambda R - kF - \omega -\mu], \\ \nonumber 
\frac{dF}{dt}&=&F[d+eR-fF-hU] .
\end{eqnarray}
Clearly, by redefining $\beta=(1-q)\lambda$ for $\omega=0$ we get the same model studied in \cite{EV95}. For the convenience of the reader
we summarize the basic results on the equilibria in which at least one of the population vanishes and then extend the study for the coexistence,
to encompass here the situation $\omega\ne 0$ not considered in \cite{EV95} for this specific equilibrium.

\subsection{Equilibria}
The equilibria are again all the equilibria of the system (\ref{mod_2_susceptibles}), namely the origin
$\widehat P_1\equiv P_1\equiv \widetilde P_1$, and
$\widehat P_2\equiv P_2\equiv \widetilde P_2$, $\widehat P_3\equiv P_3$, $\widehat P_4\equiv P_4$.
For feasibility of $\widehat P_4$ clearly we need again (\ref{P4_feas}).
Then we have
$$
\widehat P_5 = \left({\omega +\mu \over \lambda (1-q)}, \frac{(a-b\widehat R_5)\widehat R_5}{\mu}, 0 \right),
$$
which is feasible if
\begin{equation}\label{hP5_feas}
a\lambda (1-q)\ge b (\omega+\mu).
\end{equation}

Coexistence $\widehat P_6=(\widehat R_6,\widehat U_6,\widehat F_6)$ is obtained by solving the second equation in (\ref{mod_2}) at equilibrium
and substituting into the third equation of (\ref{mod_2}) to get
\begin{eqnarray*}
\widehat F_6 = {(1-q) \lambda \widehat R_5 - \omega -\mu \over k}, \quad
\widehat U_6=\left( {e \over h} - {f \over hk}(1-q) \lambda \right) \widehat R_5 + {d \over h} +
{f \over hk} (\omega +\mu),
\end{eqnarray*}
and finally from the first equation in (\ref{mod_2}) we get the quadratic
equation $\sum _{k=0}^2 c_k R^k$, whose roots determine the value of $\widehat R_6$, with
$c_0= (d k \omega + f \omega(\omega +\mu)) (hk)^{-1}>0$ and
\begin{eqnarray*}
c_2=\left( {c \over k} - {e \over h} \right) (1-q)\lambda + {f \over hk}(1-q)^2 \lambda^2-b, \\
c_1= a + {c \over k} (\omega +\mu) + {e \over h} \omega - (1-q)\lambda \left({d \over h}
+ 2{f \over hk} (\omega +\mu) \right).
\end{eqnarray*} 
Again we can apply Descartes' rule to have at least a positive root. This occurs for one root if we impose either one of the
alternative conditions
\begin{eqnarray}\label{hP6_feas0}
c_2<0, \quad c_1<0; \qquad c_2<0, \quad c_1>0,
\end{eqnarray}
and we get two positive roots if
\begin{eqnarray}\label{hP6_feas00}
c_2>0, \quad c_1<0.
\end{eqnarray}
We do not write explicitly these conditions. For feasibility we must impose
\begin{eqnarray}\label{hP6_feas1}
\widehat R_6 > \frac {\omega+\mu} {(1-q) \lambda k}
\end{eqnarray}
and the condition
\begin{eqnarray}\label{hP6_feas2}
\widehat R_6 > \frac {dk+f (\omega+\mu)} {ek - f (1-q) \lambda }, \quad
ek > f (1-q) \lambda ,
\end{eqnarray}
since the opposite one $ek < f (1-q) \lambda $ would give a negative value for $\widehat R_6$.

\subsection{Stability}

The Jacobian in this case is
$$
\widehat J= 
\left[
\begin{array}{ccc}
\widehat J_{11} & -(1-q)\lambda R+\omega & -cR \\
(1-q)\lambda U & \widehat J_{22} & -kU \\ 
 eF & -hF & \widehat J_{33}
\end{array}
\right],
$$
where
$\widehat J_{11}= a-2bR-(1-q)\lambda U-cF$, $\widehat J_{22}=(1-q)\lambda R-kF-\omega -\mu$,
$\widehat J_{33}= d+eR-hU-2fF $.

For $\widehat P_1$
the eigenvalues are $-\omega-\mu$, $d$, $a$, showing its instability.

The eigenvalues of $\widehat P_2$ are 
$-(dk+f(\omega+\mu))f^{-1}$, $-d$, $(af-cd)f^{-1}$, for which the
stability condition is (\ref{P2_stab}).
Here again comparing (\ref{P2_stab}) with  (\ref{P4_feas}) we observe the existence of a transcritical bifurcation,
for which the same conclusions, using the healthy prey invasion number (\ref{HPIN})
can be drawn as for the model with refuge for the healthy prey (\ref{mod_2_susceptibles}).

The eigenvalues of $\widehat P_3$ are
$(bd+ae)b^{-1}$, $[(1-q)\lambda a-b(\omega+\mu)]b^{-1}$, $-a$,
thus it is unstable.

For $\widehat P_4$ one eigenvalue can easily be factored out, while
the other ones are the roots of the quadratic equation (\ref{q1}). Thus, as found formerly,
by feasibility (\ref{P4_feas}) both its roots
have negative real part, and stability depends only on the first eigenvalue, namely it is given by
$(1-q)\lambda R_4<kF_4+\omega+\mu$, a condition that
can also be explicitly written as
\begin{eqnarray}\label{hP4_stab}
(1-q)\lambda\frac{af-cd}{bf+ce}<k\frac{ae+bd}{bf+ce}+\omega+\mu.
\end{eqnarray} 

An eigenvalue of $\widehat P_5$ is $d+e \widehat  R_5-h \widehat  U_5$. The other ones are the roots of
$T(\theta) = \theta^2-c_1 \theta+c_2=0$, with
\begin{eqnarray*}
c_1 = a-2b \widehat R_5-(1-q) \lambda \widehat U_5 = -\omega \frac {\widehat U_5} {\widehat R_5 } -b\widehat R_5 <0,
\quad c_2 = \mu (1-q) \lambda  \widehat U_5>0,
\end{eqnarray*}
so that both roots have negative real parts. Stability is achieved for
\begin{equation}\label{hP5_stab}
h \left( a-b {\omega +\mu \over \lambda (1-q)}\right) {\omega +\mu \over \lambda (1-q)}
> \mu \left( d+e {\omega +\mu \over \lambda (1-q)} \right) .
\end{equation}

\begin{figure}[!htb]
\centering
\includegraphics[width=8cm]{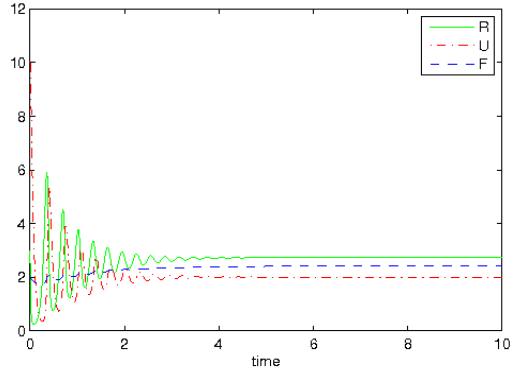}
\caption{The coexistence equilibrium $\widehat P_5$
is attained for the same parameters (\ref{param_val}) as in Figure \ref{fig-P5} with
$q=0.1$.}
\label{hatP5}
\end{figure}

Figure \ref{hatP5} shows the result of a simulation with the same parameter values (\ref{param_val}) as for Figure \ref{fig-P5}, but for $q=0.1$,
assessing the stability of the coexistence equilibrium $\widehat P_5$.

\section{Culling}

In order to eradicate the disease, another common method employed is the elimination of the infected individuals, once spotted.
Let $u(U)$ denote the control policy exercised by the farmer or the veterinarians on the infected population. We
assume it to be a linear
function of the number of infected, $u(U)=\delta U$. This control measure is of course assumed to be alternative to the safety niches.
Therefore the model (\ref{mod_2_susceptibles}), without safety niche, then modifies as follows
\begin{eqnarray}\label{mod_culling}
R'&=&R[a-bR-cF-\lambda U] + \omega U , \label{mod_1}\\ \nonumber
U'&=&U[\lambda R - kF - \omega] -(\delta +\mu) U,\label{mod_3}\\ \nonumber
F'&=&F[d+eR-fF-hU] .
\end{eqnarray}
where all the parameters retain their meaning as in (\ref{mod_2_susceptibles}).

\subsection{Equilibria}

Since only the infected equation in (\ref{mod_culling}) is affected by this change,
for $U=0$,
we easily find the very same points $\bar P_1=P_1$, $\bar P_2=P_2$, $\bar P_3=P_3$,
$\bar P_4=P_4$ of (\ref{mod_2_susceptibles}), the latter having clearly the same feasibility condition
(\ref{P4_feas}).

For $U > 0$ we find the predator-free point
$$
\bar P_5=\left(\frac{\omega+\delta+\mu}{\lambda}, \frac{a\lambda(\omega+\delta+\mu)-b(\omega+\delta+\mu)^2}{(\delta+\mu)\lambda^2},0\right),
$$
It is feasible for
\begin{equation}\label{Q5_feas}
a\lambda \ge b (\omega + \delta+\mu).
\end{equation}

We then have the coexistence equilibrium $\bar P_6=[\bar R_6,\bar U_6,\bar F_6]$, whose population values are found by solving the last equilibrium
equation in (\ref{mod_culling}) to get
\begin{eqnarray*}
\bar F_6 = { \lambda \bar R_6 - \omega-\delta -\mu\over k},
\end{eqnarray*}
and by substituting into the second one of (\ref{mod_culling}) we find
\begin{eqnarray*}
\bar U_6 =\left( {e \over h} - {f \over hk}\lambda \right) \bar R_5 + {d \over h} + {f \over hk}
(\omega + \delta+\mu),
\end{eqnarray*}
and finally from the first one of (\ref{mod_culling}) we get the quadratic equation $a_2 R^2+a_1R+a_0=0$ with
\begin{eqnarray*}
a_2=-bhk -\left( ch - ek \right) \lambda + f \lambda^2, \quad
a_0= d k \omega + f \omega (\omega+\delta+\mu), \\
a_1= (ak + c (\omega + \delta+\mu) - \lambda k) \left( d k + f (\omega+\delta+\mu) \right)
+ \omega \left( ek -f\lambda \right),
\end{eqnarray*}
whose positive roots give the value of $\bar R_6$.
Since $a_0>0$, imposing $a_2<0$ ensures that exactly one positive root exists. Therefore a sufficient condition for
feasibility and uniqueness is
\begin{equation}\label{Q6_feas_1}
f \lambda^2< bhk +\left( ch - ek \right) \lambda.
\end{equation}
Alternatively, there will be two positive roots if $a_1^2>4a_2a_0$, $a_2>0$ and $a_1<0$, a situation that we however do not explore any further.

For feasibility, we need further to require
\begin{equation}\label{Q6_feas_2}
\bar R_6 > \frac 1{\lambda} ( \omega+\delta +\mu), \quad (f\lambda -ek) \bar R_6 < kd+f ( \omega+\delta +\mu).
\end{equation}

\subsection{Stability}
The Jacobian of (\ref{mod_culling}) is
$$
\bar J= 
\left[
\begin{array}{ccc}
a-2bR-\lambda U-cF & -\lambda R+\omega & -cR \\
\lambda U & \lambda R-kF-\omega-\delta -\mu& -kU \\ 
 eF & -hF & d+eR-hU-2fF 
\end{array}
\right].
$$

Minor modifications involve the eigenvalues at the equilibria that coincide with those of (\ref{mod_2_susceptibles}).
$\bar P_1$ and $\bar P_3$ retain their instability here too,
$\bar P_2$ is stable when (\ref{P2_stab}) holds, $\bar P_4$ has two eigenvalues with
negative real parts as for (\ref{mod_2_susceptibles}), but the first one now also contains the culling term, so that the stability
condition (\ref{P4_stab}) gets here replaced by the more general condition
$\lambda \bar R_4<k \bar F_4+\omega+\delta+\mu$ or, explicitly,
\begin{equation}\label{cP4_stab}
\lambda (af-cd) < k(bd+ae) + (ce+bf)(\omega+\delta+\mu).
\end{equation}

For $\bar P_5$ one eigenvalue is $d+e \bar R_5-h \bar U_5$.
The other ones are the roots of $T(\theta) = \theta^2+c_1 \theta+c_2=0$, with
\begin{eqnarray*}
c_1 = a-2b\bar R_5-\lambda \bar U_5, \quad c_2 = \delta\lambda \bar U_5>0.
\end{eqnarray*}
Explicitly,
$$
T_{1,2}=\frac{-c_1\pm\sqrt{c_1^2-4c_2}}{2}.
$$
By Descartes' rule of signs, both have negative real parts if we impose
$$
a-2b\bar R_5-\lambda \bar U_5<0,
$$
i.e.
$$
\bar U_5>\frac{a-2b \bar R_5}{\lambda}.
$$
This inequality is always satisfied, since
using the equilibrium values, the right hand side becomes
\begin{eqnarray*}
\frac{a-2b \bar R_5}{\lambda}=\frac{a}{\lambda}-2\frac{b}{\lambda}\frac{\omega+\delta+\mu}{\lambda}
=\frac{a\lambda- 2b (\omega + \delta+\mu)}{\lambda^2}\\
=\frac{\delta \bar U_5}{\omega+\delta+\mu}- \frac{b(\omega+\delta+\mu)}{\delta^2}
< \bar U_5-\frac{b(\omega+\delta+\mu)}{\delta^2},
\end{eqnarray*}
and the last expression is always smaller than $\bar U_5$ as required.
Stability hinges on the first eigenvalue only, giving
\begin{equation}\label{Q5_stab}
\bar U_5> \frac{d+e \bar R_5 }{h}.
\end{equation}

\section{Bifurcations}

In this short Section we highlight a few other features of the models.

For the model (\ref{mod_2_susceptibles}), i.e. the refuge for the healthy prey, there is a transcritical bifurcation for
which $P_4$ emanates from $P_2$ when the parameters satisfy the critical condition
\begin{equation}\label{TC}
af=cd,
\end{equation}
compare (\ref{P2_stab}) and (\ref{P4_feas}).

Furthermore $P_2$ and $P_5$ are both simultaneously stable if
both (\ref{P2_stab}) and (\ref{P5_stab}) hold. Rewriting extensively the latter, we find indeed that (\ref{P5_feas})
is its consequence. Explicitly, we have
$$
af<cd,\quad \frac{h(a\lambda( \lambda s +\omega +\mu) -b( \lambda s +\omega +\mu)^2)}{\lambda^2 \mu}
>d+\frac{e(\lambda s +\omega + \mu)}{\lambda}.
$$
Also $P_4$ and $P_5$ are stable simultaneously if
$$
af>cd,\quad \frac{h(a\lambda( \lambda s +\omega +\mu) -b( \lambda s +\omega +\mu)^2)}{\lambda^2 \mu}
>d+\frac{e(\lambda s +\omega + \mu)}{\lambda}.
$$

In case of the reduced contacts, model (\ref{mod_2}), bistability occurs
between the same two pairs of equilibria, with slightly different conditions, namely
$$
af<cd,\quad \frac{h(a\lambda(1-q)( \omega +\mu) -b( \omega +\mu)^2)}{\lambda^2 (1-q)^2 \mu}>d+\frac{e(\omega + \mu)}{\lambda(1-q)}.
$$
for $\widehat P_2$ and $\widehat P_5$, while for $\widehat P_4$ and $\widehat P_5$ they become
$$
af>cd,\quad \frac{h(a\lambda(1-q)( \omega +\mu) -b( \omega +\mu)^2)}{\lambda^2 (1-q)^2 \mu}>d+\frac{e(\omega + \mu)}{\lambda(1-q)}.
$$

\begin{figure}[!htbp]
\centering
\includegraphics[width=5.8cm]{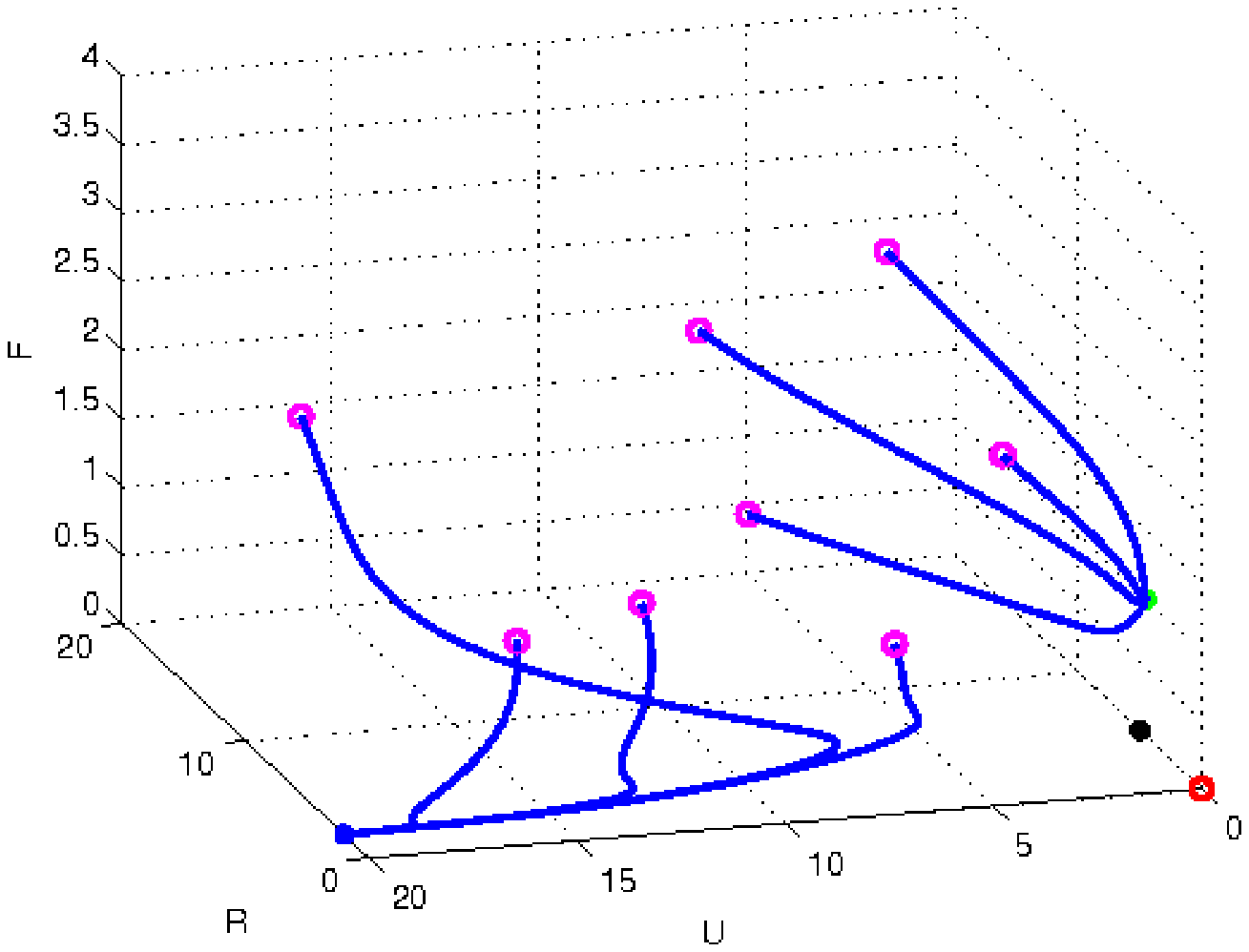}  
\includegraphics[width=5.8cm]{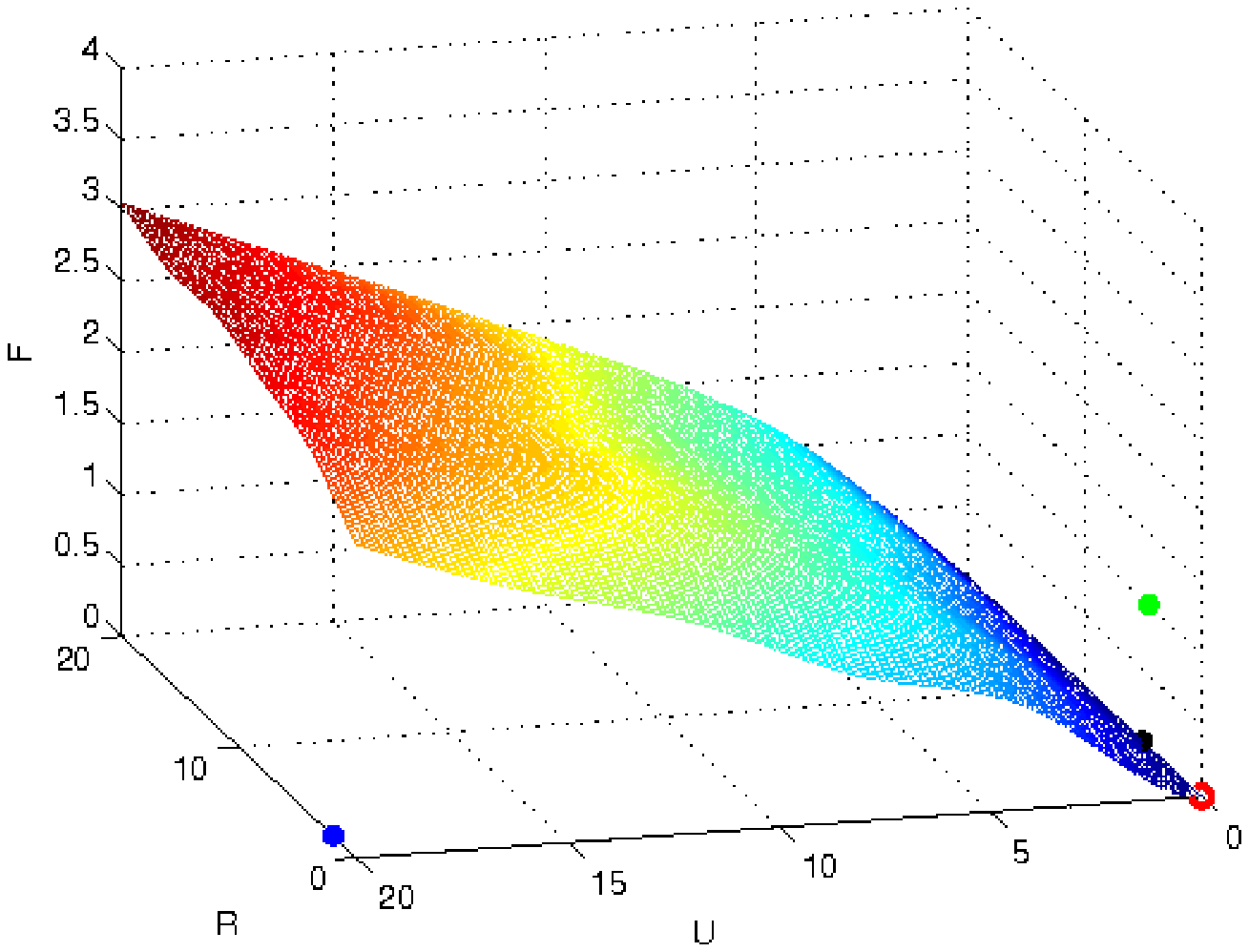}  
\caption{Bistability of $P_4$ and $\bar P_5$. Left frame: trajectories from
different initial conditions tend to the two equilibria; Right: separatrix surface. Both
plots are obtained for the following set of parameter values
$ a=10, b=2, c=1, d=0.1, e=0.2, f=1, h=0.2, k=3, \lambda=0.75, \omega=0.9, \delta=0.6$.
Note that the figure has been rotated, the origin lies in the bottom right corner; it is
an unstable equilibrium marked with the red dot. Also shown with a black dot on the axis is the saddle
point $\bar P_3$.}
\label{fig:abbattimento_globale}
\end{figure}

Finally, for the model with culling, (\ref{mod_culling}), the very same pairs of points are providing
bistability once again, with conditions
for $\bar P_2$ and $\bar P_5$ given by
$$
af<cd, \quad
\frac{a\lambda(\omega+\delta+\mu)-b(\omega+\delta+\mu)^2}{(\delta+\mu)\lambda}
> \frac{d\lambda+ e (\omega+\delta+\mu)}{h},
$$
while those for $\bar P_4$ and $\bar P_5$ are
$$
af>cd, \quad
\frac{a\lambda(\omega+\delta+\mu)-b(\omega+\delta+\mu)^2}{(\delta+\mu)\lambda}
> \frac{d\lambda+ e (\omega+\delta+\mu)}{h}.
$$
This result is illustrated in Figure \ref{fig:abbattimento_globale},
for the parameter set
$ a=10, b=2, c=1, d=0.1, e=0.2, f=1, h=0.2, k=3, \lambda=0.75, \omega=0.9, \delta=0.6$. The points $\bar P_1$,
$\bar P_2$, $\bar P_3$, $\bar P_4$ and $\bar P_5$ are all feasible.
The equilibria $\bar P_4$ and $\bar P_5$
are both stable, while $\bar P_1$, $\bar P_2$, $\bar P_3$ are not.

\section{Models Comparison}\label{comparison}

\subsection{The underlying demographic model}
The classical quadratic predator-prey model underlying these ecoepidemic systems is obtained by
eliminating the variable $U$ and its corresponding equation in (\ref{mod_2}). This differs
from the classical Lotka-Volterra model in which no extra food source is available for predators,
which therefore experience an exponential mortality in absence of the prey. A related model in which
the predator's carrying capacity depends on the prey population size had been introduced in \cite{M73}.
The reduced system with no infected,
which is the projection of (\ref{mod_2}) onto the disease-free $R-F$ phase plane, has the following equilibria:
\begin{eqnarray*}
Q_1=(0,0), \quad
Q_2=\left( 0,\frac{d}{f}\right), \quad
Q_3=\left( \frac{a}{b},0\right), \quad
Q_4=\left( \frac{af-cd}{bf+ce},\frac{ae+bd}{bf+ce}\right).
\end{eqnarray*}
The latter is feasible when (\ref{P4_feas}) holds.

$Q_1$ and $Q_3$ are both unstable, in view of their respective eigenvalues
$a$, $d$ and
$-a$, $(ae+bd)b^{-1}$.
For $Q_2$ we find 
$(af-cd)f^{-1}$, $-d$
showing that it is stable exactly when (\ref{P2_stab}) holds.
The eigenvalues of $Q_4$ are complex conjugate, with negative real part, so that $Q_4$ is unconditionally stable. Being the only such equilibrium,
local stability implies global stability. This fact could be shown also via a suitable Lyapunov function.

\subsection{Models equilibria summary}
The demographic equilibria $P_1$-$P_4$ (labeled with the notation of the model (\ref{mod_2_susceptibles}))
are the same in the four systems. Of these, the first three are
always feasible, and $P_1$ and $P_3$ are always unstable. The feasibility condition for $P_4$ is always
(\ref{P4_feas}). The predator-free equilibrium differs in each case, because of the prey levels $R_5$ attained in
each model. The predators settle always at the level $\mu^{-1}R_5(a-bR_5)$. Below, we summarize the feasibility
conditions in each case.

\vspace{0.3cm}
\begin{center}
\begin{tabular}{c|cccc}
\hline
Feasibility & Model (\ref{mod_2_susceptibles}) & Model (\ref{mod_2_infected}) & Model (\ref{mod_2}) & Model (\ref{mod_culling}) \\
\hline
$P_4$ & $af\ge cd$ & $af\ge cd$ & $af\ge cd$ & $af\ge cd$ \\
$P_5$ & $\frac ab\ge s+ \frac {\omega + \mu} {\lambda}$ & $\frac ab \ge \widetilde R_5$ &
$\frac ab (1-q) \ge \frac {\omega+\mu}{\lambda}$
& $\frac ab \ge \frac {\omega + \delta+\mu}{\lambda}$\\
\hline
\end{tabular}
\end{center}
\vspace{0.3cm}

The stability condition for $P_2$ is always
(\ref{P2_stab}). 
The stability conditions for each equilibrium, assuming feasibility, are instead

\vspace{0.3cm}
\begin{center}
\begin{tabular}{c|cccc}
\hline
Stability & Model (\ref{mod_2_susceptibles}) & Model (\ref{mod_2_infected}) &
Model (\ref{mod_2}) & Model (\ref{mod_culling}) \\
\hline
$P_2$ & $af<cd$ & $af<cd$ & $af<cd$ & $af<cd$ \\
$P_4$ & (\ref{P4_stab}) & always (when feasible) & (\ref{hP4_stab}) & (\ref{cP4_stab}) \\
$P_5$ & (\ref{P5_stab}) & (\ref{tP5_stab1}) (\ref{tP5_stab2}) & (\ref{hP5_stab}) & (\ref{Q5_stab}) \\
\hline
\end{tabular}
\end{center}
\vspace{0.3cm}

\subsection{Attainable equilibria with vanishing populations}
The ecoepidemic system exhibits a similar
range of behaviors as the demographic ecosystem: predator and prey coexistence is allowed,
both with and without infected, compare $P_4$ and $P_6$, and also
the predators-only equilibrium $P_2$; this is biologically meaningful
recalling that they have other food sources available.
Comparing feasibility and stability conditions for $P_2$ and $P_4$ a transcritical bifurcation
is seen to arise whenever (\ref{TC}) holds. This clearly stems from the purely demographic model underlying
all these ecoepidemic models.

Evidently, in the prey-free environment expressed by equilibrium $P_2$,
the role of the refuge for the prey is nonexistent. In fact the refuge-related parameters appear
neither in its feasibility nor in its stability conditions.

The same does not occur, not surprisingly either,
for the disease-free
equilibrium $P_4$. In fact its feasibility and
the population levels are not affected by the size of the refuges in any model, but the stability of this equilibrium
does in fact depend on this parameter. The way in which the refuges' parameters $s$, $q$ and $\delta$ appear in the
stability conditions differs. Considering (\ref{P4_stab}), (\ref{hP4_stab}) and (\ref{cP4_stab}),
we find that
$s$, $q$ and $\delta$ have a stabilizing effect for the ecoepidemic system,
a result which as mentioned agrees with former findings in the literature for predator-prey models, \cite{Gonzalez2012}.
In fact, in the case of the reduced contacts model, the refuge favors stability since,
mathematically, the left hand side becomes smaller due to a positive $q$,
while in the case of a refuge for the healthy prey and
of culling it is the right hand side that gets increased by the presence of $s$ and $\delta$ respectively.
However, since $q$ is a fraction, denoting the relative reduction in the
frequency of contacts, while $s$ represents the number of refuges
and $\delta$ the culling rate, it is more likely that $s$ and $\delta$ could be sensibly larger than $q$
and therefore have a more marked influence on the stability of this equilibrium. 
Comparison of (\ref{cP4_stab}) with (\ref{P4_stab}) shows that $\delta$
must be compared with $s\lambda$, to assess which model provides the
less stringent stability conditions.
Comparison of (\ref{cP4_stab}) with (\ref{hP4_stab}) instead reduces to comparing $\delta$ with
$q\lambda (af-cd)(bf+ce)^{-1}$. Similary we must compare
$s$ with $q (af-cd)(bf+ce)^{-1}$ to assess the largest stability condition between
(\ref{P4_stab}) and (\ref{hP4_stab}).


The feasibility conditions for the equilibrium $P_5$ in all models,
namely (\ref{P5_feas}), (\ref{hP5_feas}), (\ref{Q5_feas}), 
are always an explicit restatement of (\ref{tP5_feas}).
Thus the predator-free equilibrium $P_5$ entails that 
the size of surviving healthy individuals drops below the level of equilibrium $P_3$, when they
would thrive alone in the disease-free environment, if the equilibrium were stable.
This is at first sight a somewhat counterintuitive result. Indeed
it is true that the niches help the infected not to get in contact with the susceptibles,
but then one would expect also an advantage for the
healthy individuals. On the other hand, we can explain it saying that they cannot exceed
the carrying capacity of the environment, which is exactly achieved by the healthy prey
when they would thrive alone in the predator-free environment, at $P_3$.
While the presence of the predators could contribute toward the eradication of the disease helping the system
to settle at $P_4$,
their absence cannot improve the environment conditions so that the healthy prey grow beyond the level allowed
by the available resources.
Another way of looking at this situation is to observe that in this case the niche stabilizes
the otherwise unstable predator-free equilibrium, at the price of making the disease endemic.

When feasible, the predator-free equilibrium $P_5$ is stable if the conditions 
(\ref{P5_stab}), (\ref{tP5_stab1}), (\ref{hP5_stab}), (\ref{Q5_stab}) hold, all expressing the same
relation, while the model with the cover for infected in addition needs also (\ref{tP5_stab2}). To
compare effects of the various types of refuge is not immediate. The refuge-related
parameters appear in all models in both healthy and infected prey, and therefore on both sides of the
stability conditions. The latter are reduced to the inequality
$$
bh R_5^2 +(e\mu-ah) R_5 +d\mu <0,
$$
in which $R_5$ has the value provided by each model. Denoting by $R_5^{\pm}$ the roots of the associated
equation to the above inequality, which are real if
\begin{equation}\label{discr}
(ah-e\mu)>4bdh\mu,
\end{equation}
a condition that we now assume, the effect of the cover in each case can be estimated via the
inequalities
$$
R_5^- \le s + \frac {\omega+\mu}{\lambda} \le R_5^+, \quad 
R_5^- \le \frac {\omega+\mu}{(1-q)\lambda} \le R_5^+, \quad 
R_5^- \le \frac {\omega+\mu+\delta}{\lambda} \le R_5^+.
$$

\subsection{Models coexistence equilibria}
The numerical experiments with the  coexistence equilibria of the three models show that
using the set of demographic parameter values in (\ref{param_val}), i.e. those given by the first row,
the system settles to the demographic disease-free equilibrium
$(23.2475, 0, 14.0261)$, whose projection onto the $R-F$ phase plane corresponds of course to the equilibrium of the underlying
classical predator-prey system,
$(23.2475, 14.0261)$. If we now introduce the disease, with the related parameter values found in the second row of (\ref{param_val}),
we find the ecoepidemic equilibrium
$(2.7450, 1.7848, 2.4334)$. As we can easily observe, the disease has a large impact on the system,
reducing both its populations by an order of magnitude. Although the epidemics affects only the prey,
its effect is felt also by the predators. This can easily be interpreted, because a reduced food supply,
due to a lower prey population caused by the disease, must reduce also the predator population and, in addition, consumption of infected
prey is harmful for the predators. In other words,
diseases, as stated many times in ecoepidemiological research, affect the whole ecosystems, and therefore in environmental studies they cannot
be easily neglected.

\subsection{Effects of safety refuges on coexistence}
\begin{figure}[!htb]
\centering
\includegraphics[width=5.8cm]{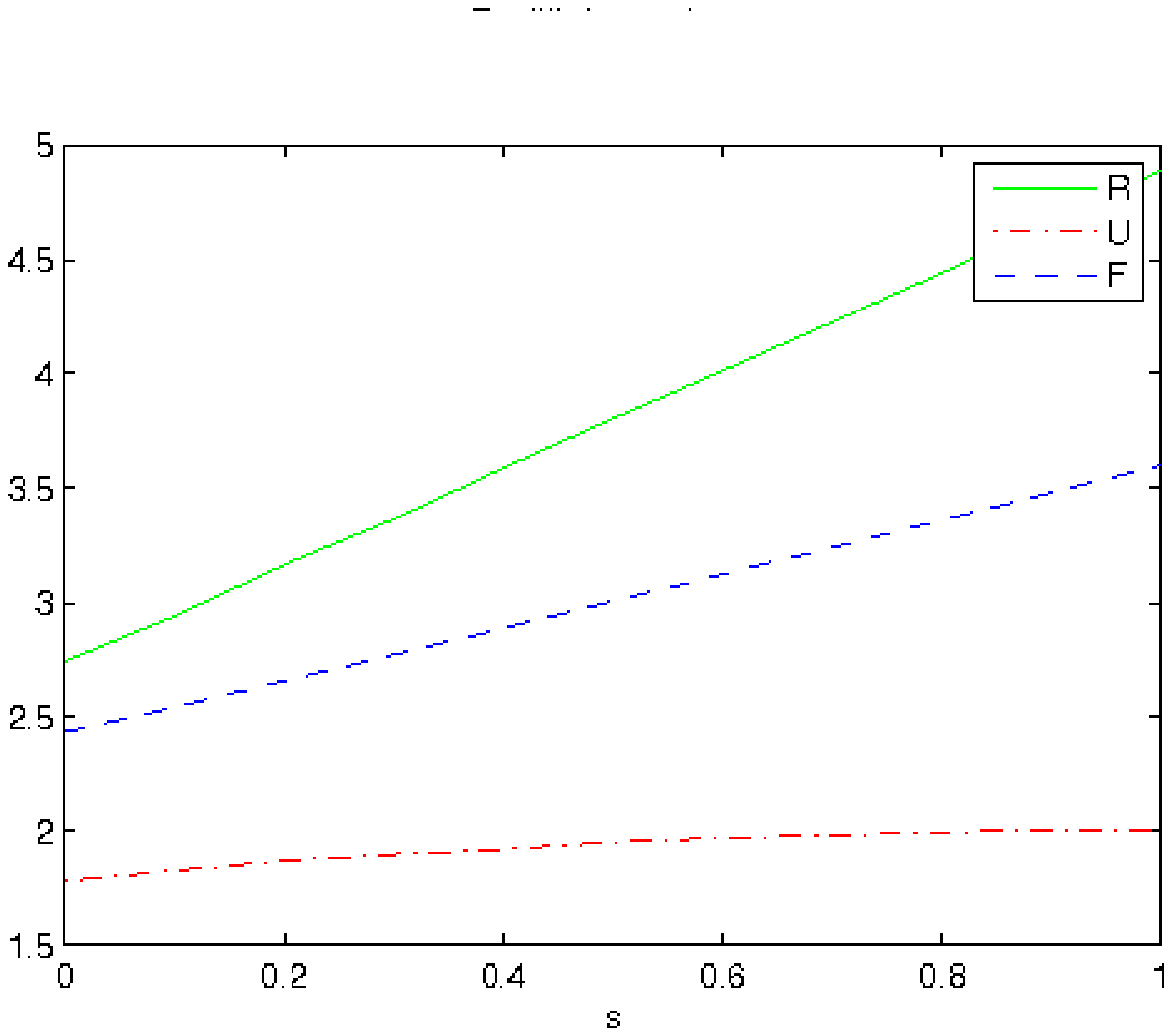}
\includegraphics[width=5.8cm]{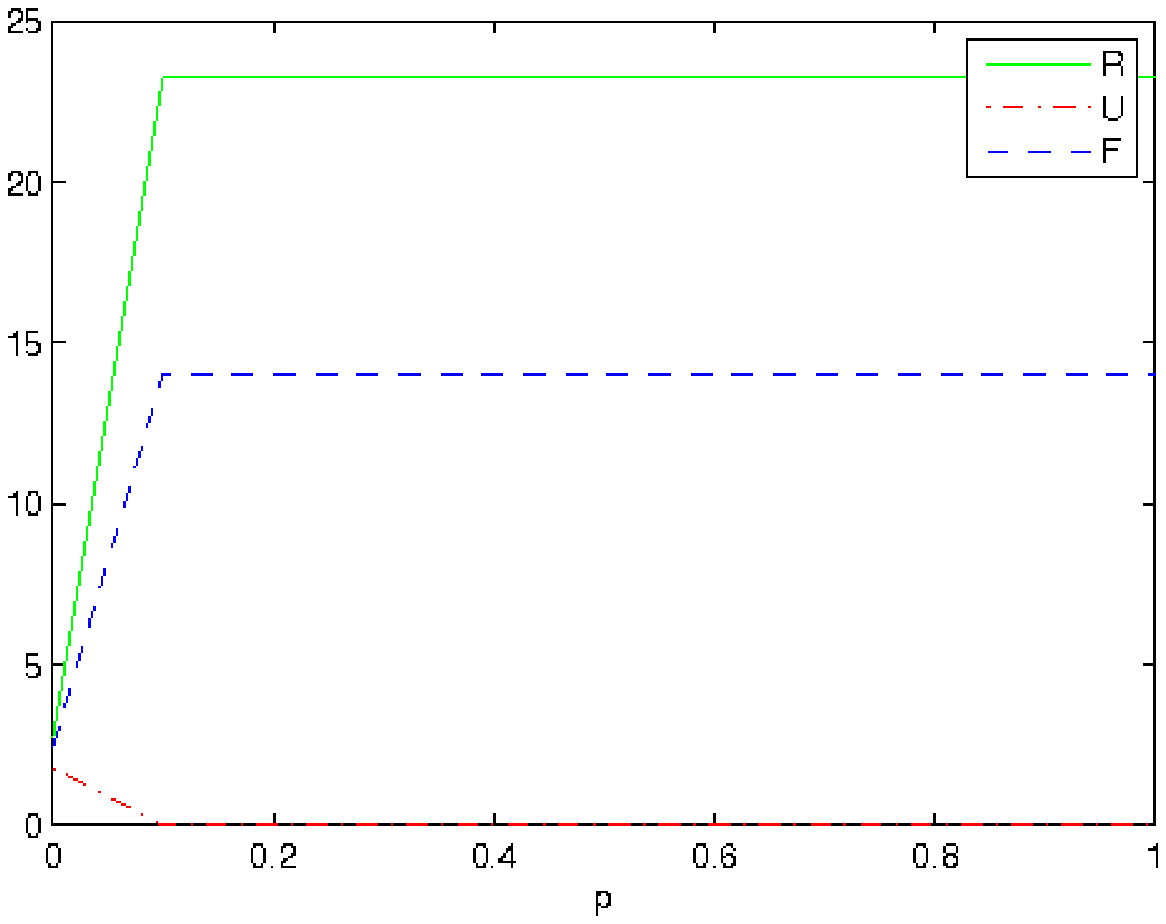}\\
\includegraphics[width=5.8cm]{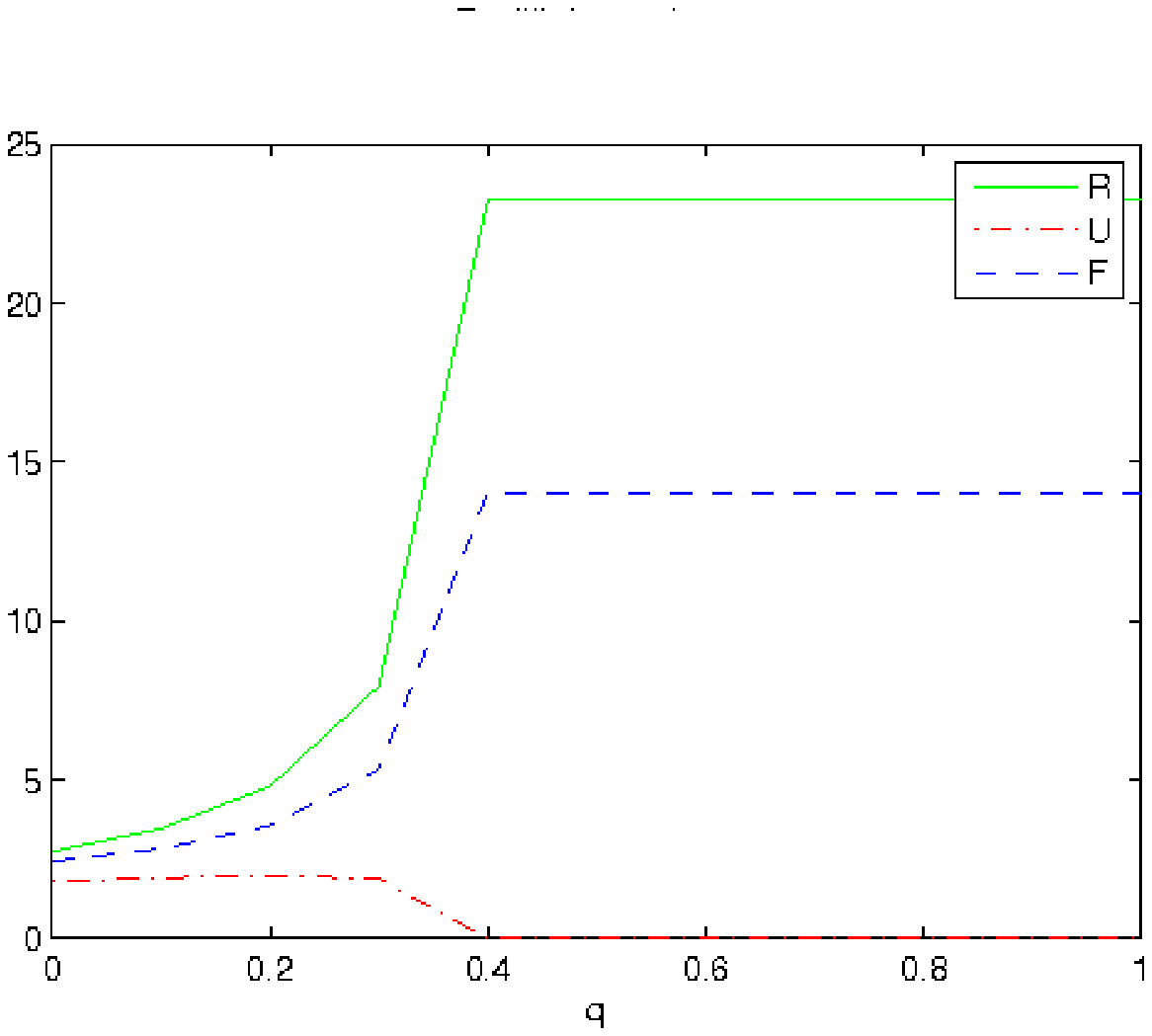}
\includegraphics[width=5.8cm]{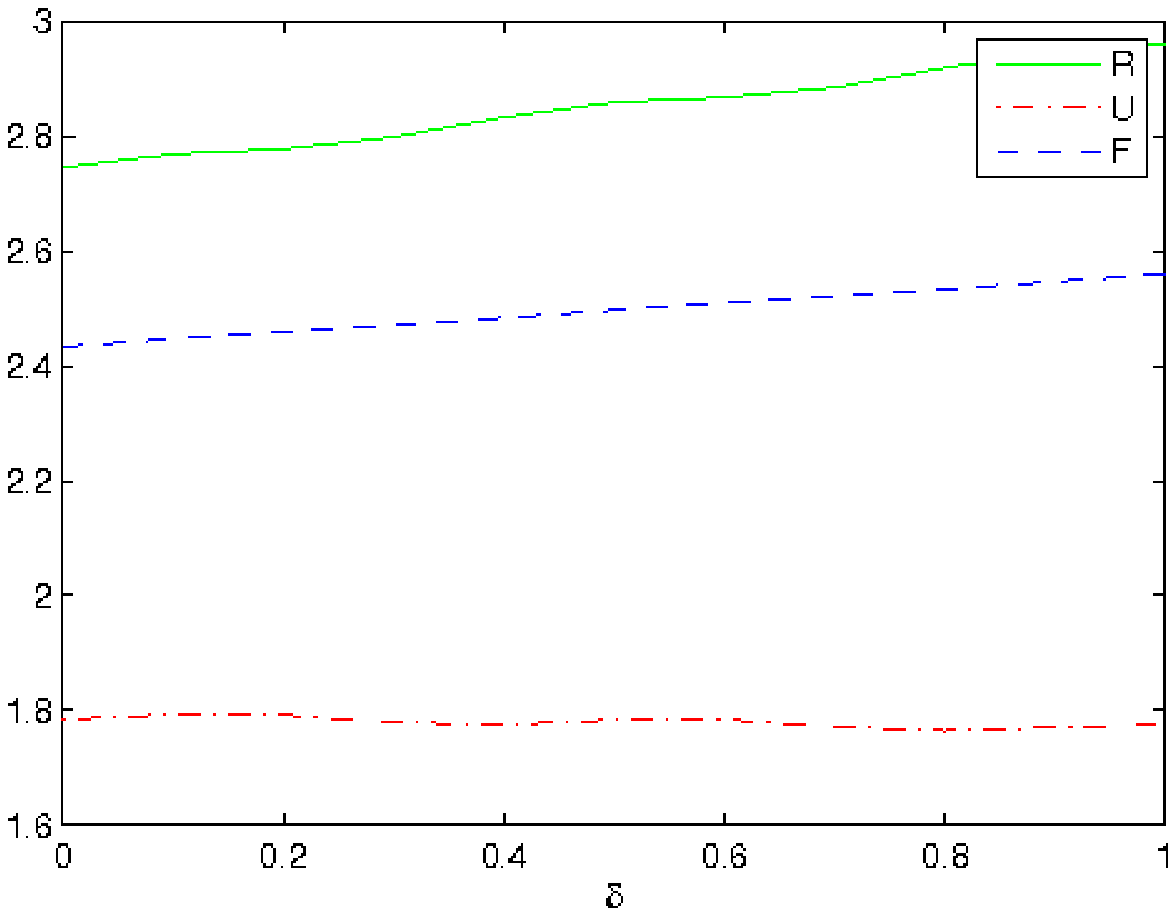}
\caption{Equilibrium population values of system (\ref{mod_2_susceptibles}) as function of the controls. Clockwise from the
upper left corner: refuge size $s$, refuge size $p$, culling rate $\delta$, refuge size $q$.}
\label{niches}
\end{figure}

Coming back to the effects of our safety refuges, we have run simulations using the
previous parameter values (\ref{param_val}), with various sizes for the refuge coefficients $s$, $p$ and $q$. As remarked earlier
the proviso holds, that in the models (\ref{mod_2_susceptibles}) and (\ref{mod_2_infected})
a check is implemented, for which when $U<p$ and $R<s$ the next to last term in
the first equation and the first one in the second equation are set to zero
in both (\ref{mod_2_susceptibles}) and (\ref{mod_2_infected}). The results are reported in Figure \ref{niches}.

Comparison of the results indicates that for the healthy refuge, the healthy prey and the predators at equilibrium increase in
a linear fashion their numbers as $s$ grows, while the infected appear to reach a plateau. When the infected prey have a cover,
there is a threshold value of its size $p$ beyond which the disease disappears and the other populations suddenly jump to the
level of the corresponding demographic, disease-free, classical model and stay there independently of the value of $p$.
A similar result holds also when it is the contact rate that gets reduced, i.e. for model (\ref{mod_2}).
In this case the equilibria behavior before the threshold value
of $q$ is reached appears to be smoother than in the previous case of system (\ref{mod_2_infected}).
For the culling policy instead, in this case at least,
the healthy prey slightly increase their levels as the rate $\delta$ grows,
but the infected do not vary much and in particular the disease is not eradicated.

\begin{figure}[!htb]
\centering
\includegraphics[width=10cm]{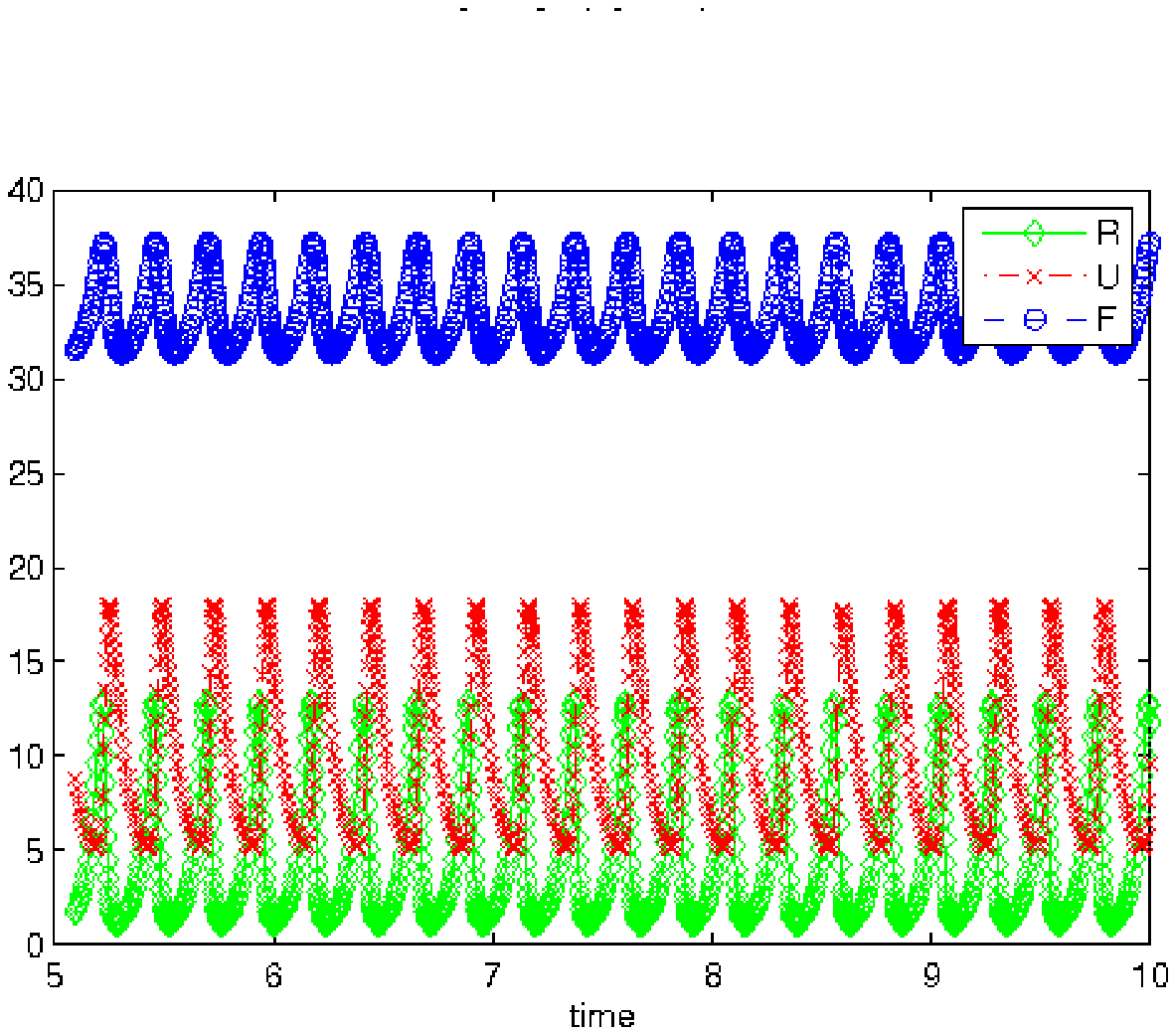}
\caption{Limit cycles obtained when the control over the infected is exercised through a refuge.}
\label{fig:oscill}
\end{figure}

We also discovered persistent oscillations triggered by the use of infected refuges, i.e. through the parameter $p$, Figure \ref{fig:oscill}.

\subsection{Combined effects of refuge and epidemiological parameters}

\begin{figure}[!htb]
\centering
\includegraphics[width=8cm]{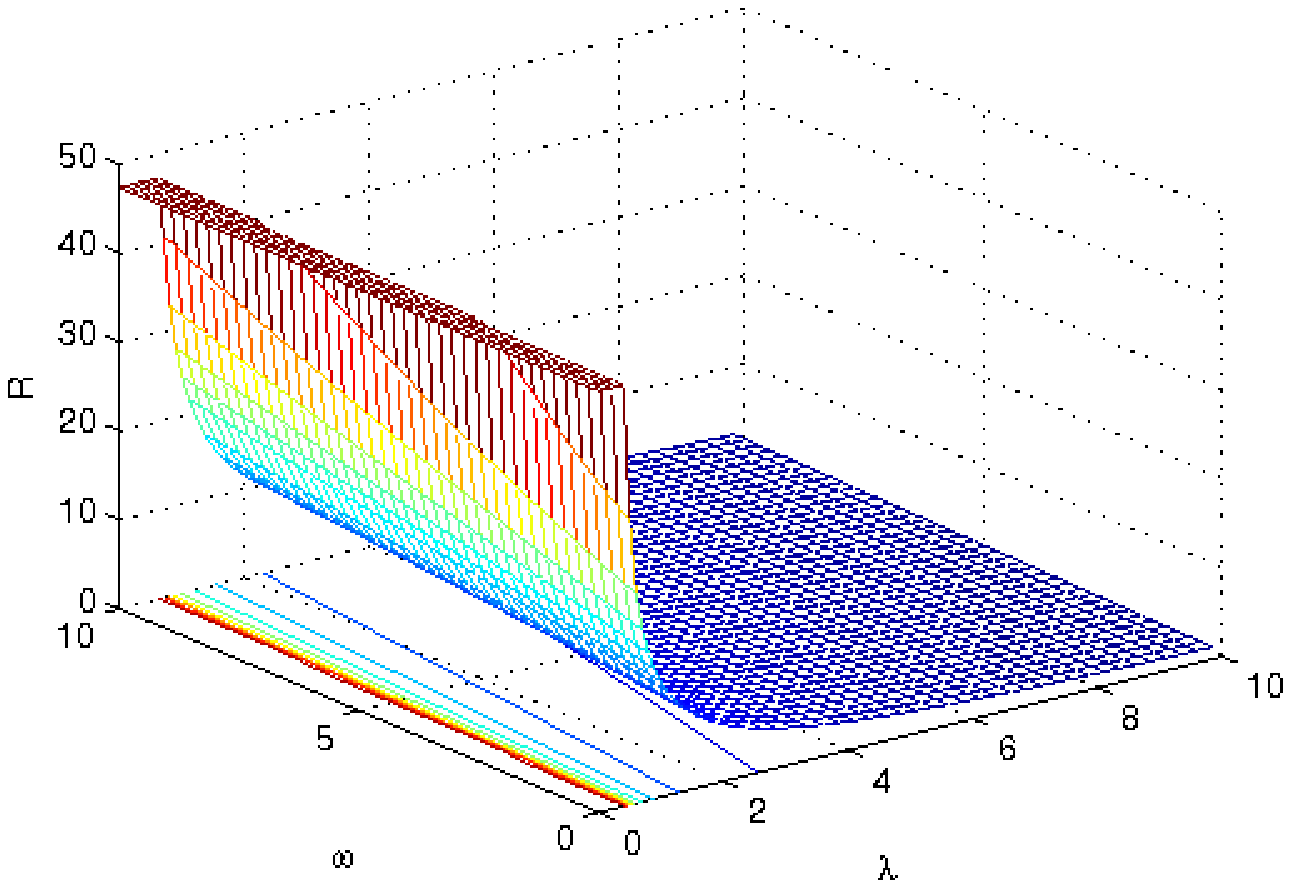}
\includegraphics[width=8cm]{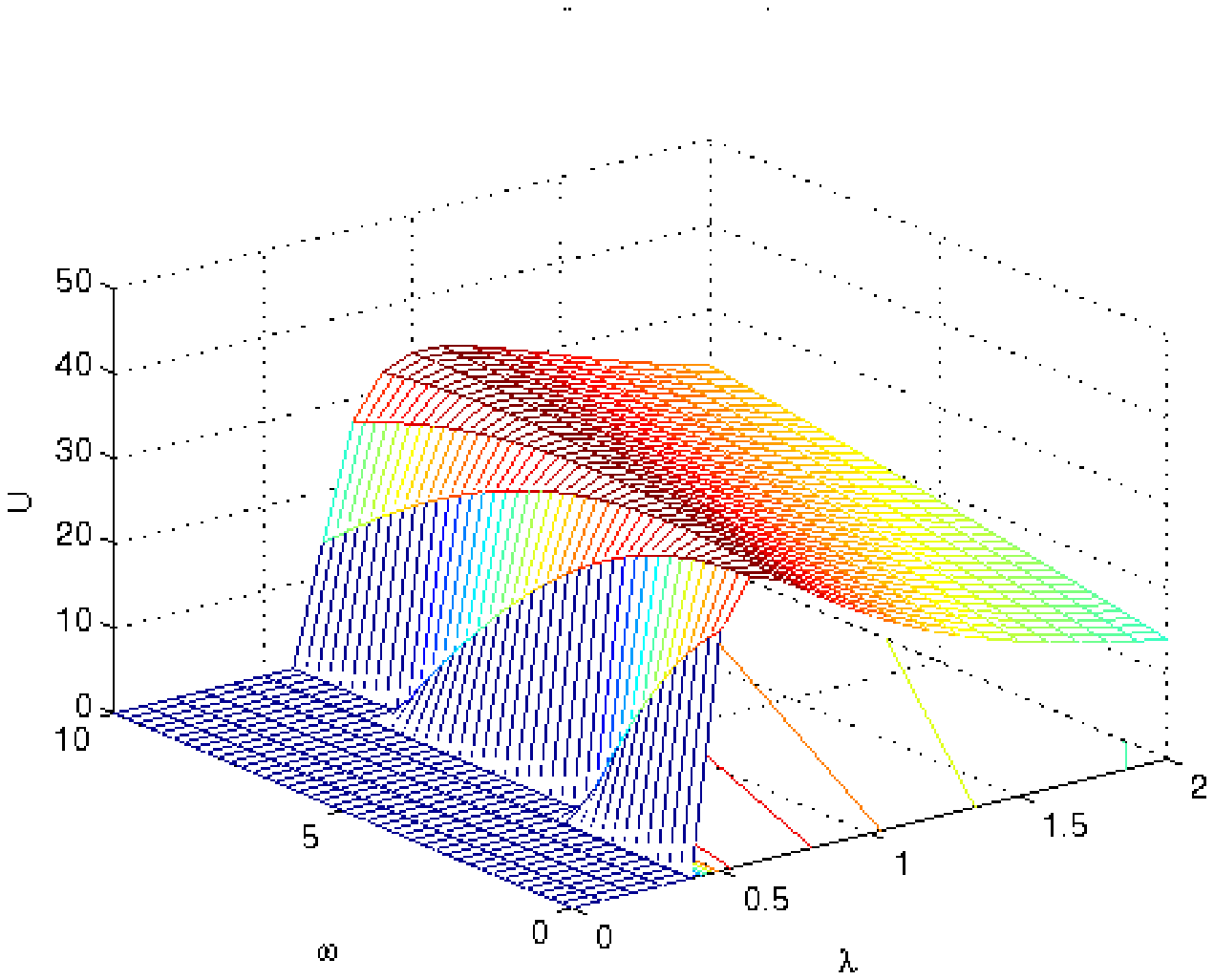}
\includegraphics[width=8cm]{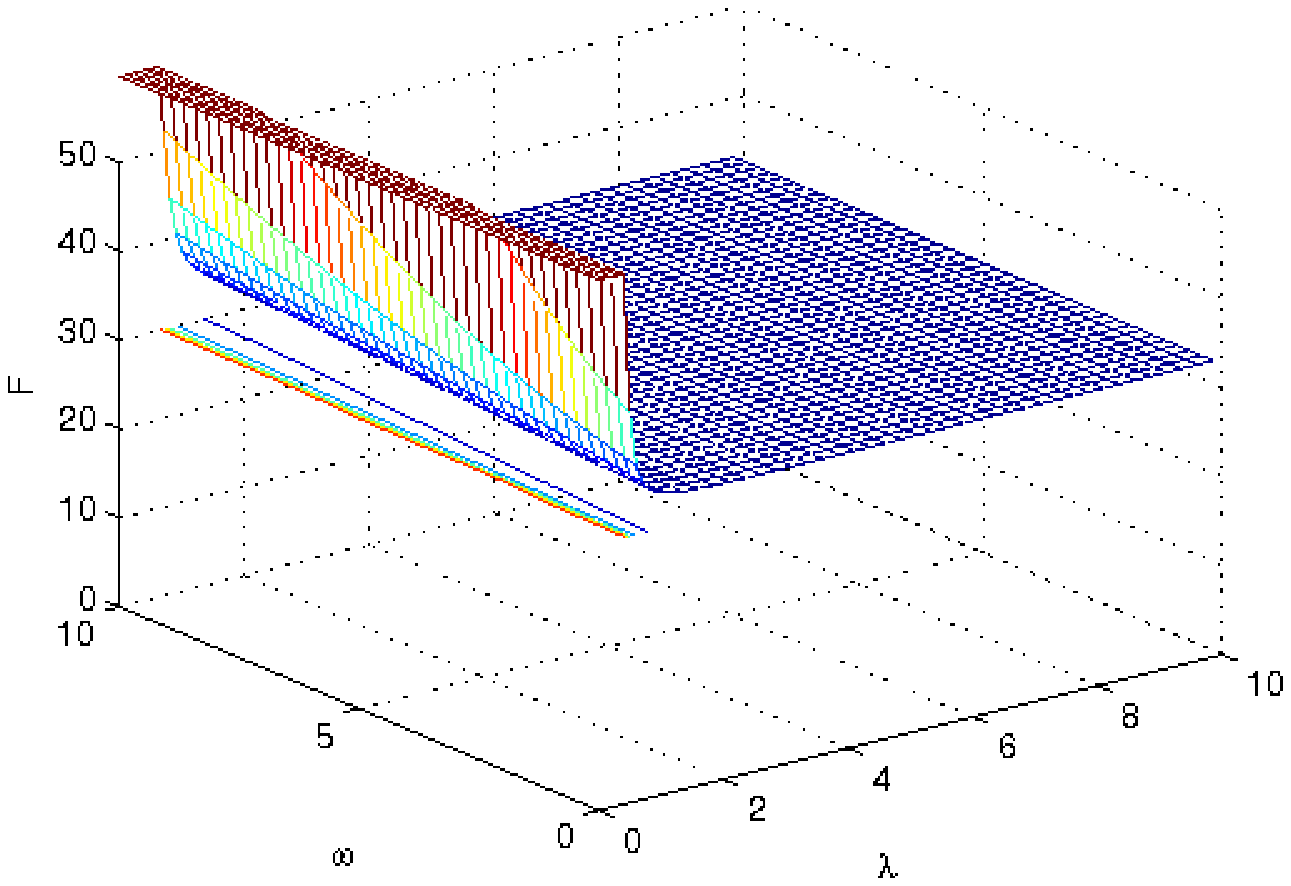}
\caption{No disease control: top healthy prey; center infected prey; bottom predators.}
\label{fig:no_contr}
\end{figure}

Consider the ecoepidemic model without any disease control. In Figure \ref{fig:no_contr} center we show the infected level as a function of
the epidemic parameters $\lambda$ and $\omega$ for a fixed choice of the demographic parameters, namely
\begin{equation}\label{param_set_2}
a=50, \quad b=0.3, \quad d=30, \quad e=0.5, \quad c=0.6, \quad f=0.9, \quad h=0.23, \quad k=0.3.
\end{equation}

Observing the level of infected, we choose as reference values for the epidemic parameters $\lambda=0.8$ and $\omega=5$, corresponding
to the peak in the infectives. When performing simulations with the various disease controls,
we will show the simulations results versus the control parameter and one of the epidemic parameters at the time.
When using as epidemic parameter $\lambda$ for instance, we will have to compare the figure with the
line in Figure \ref{fig:no_contr} given by the intersection of
the surface with the plane $\omega=5$. This function raises up to a maximum
and then decreases. This function has to be compared with the situation when some control is implemented.

\begin{figure}[!htb]
\centering
\includegraphics[width=8cm]{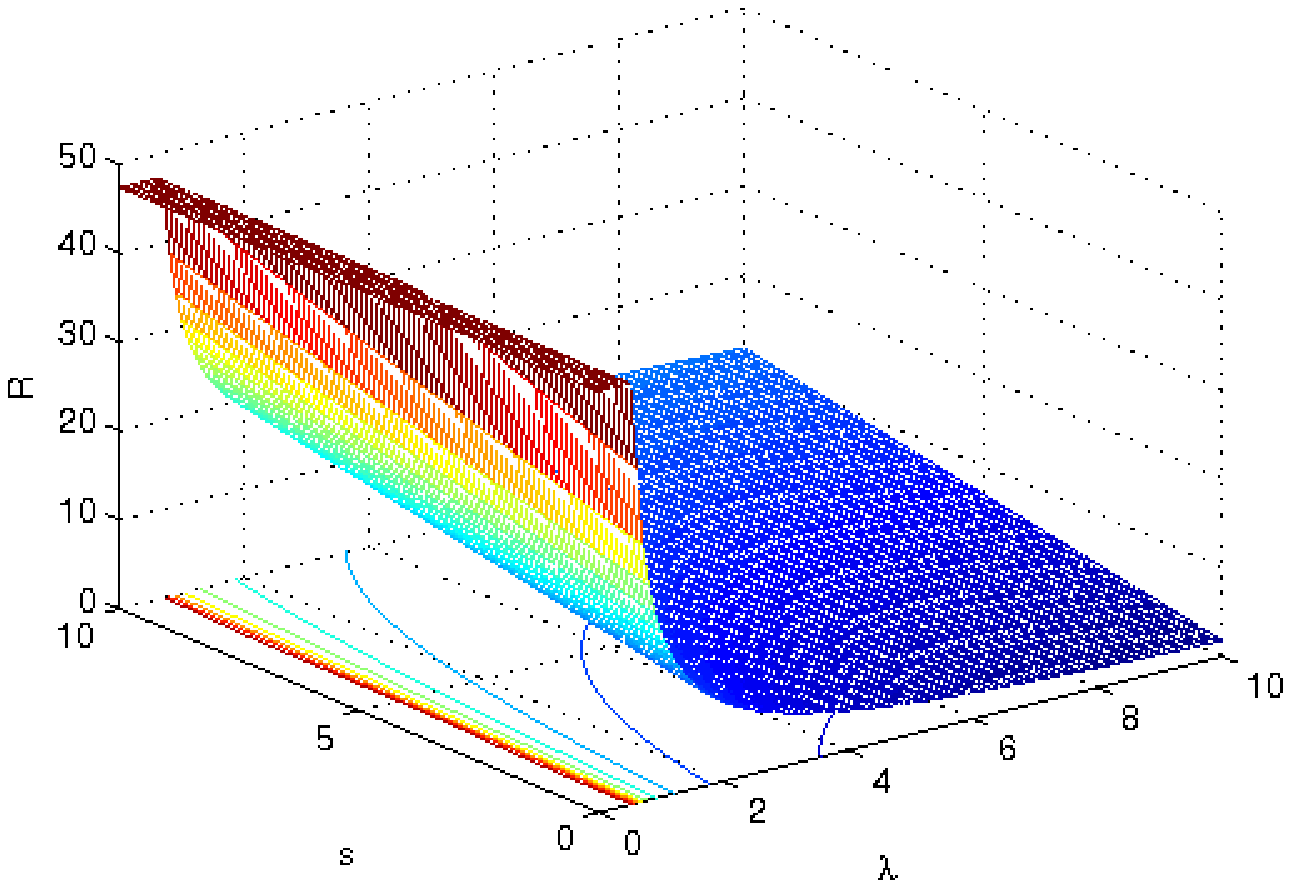}
\includegraphics[width=8cm]{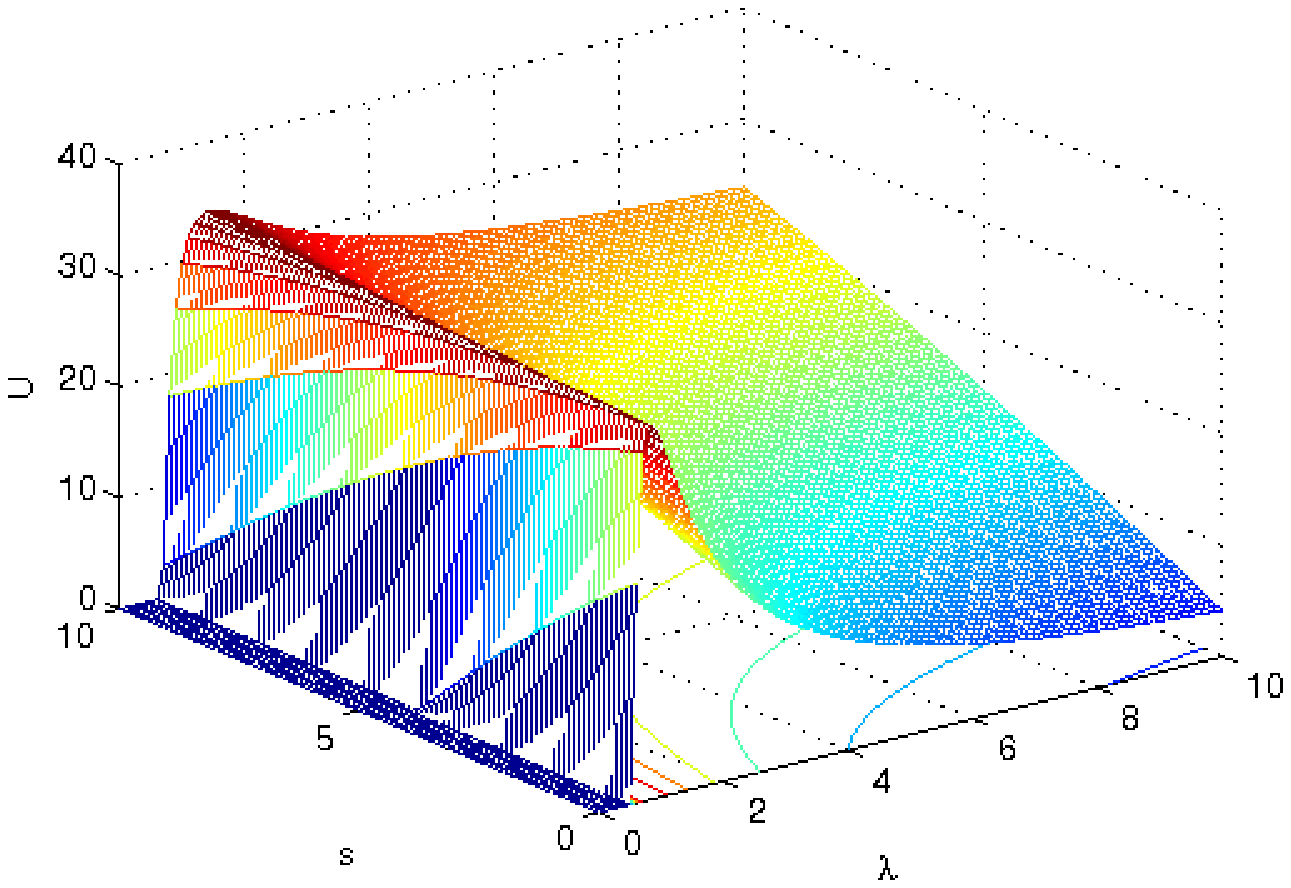}
\includegraphics[width=8cm]{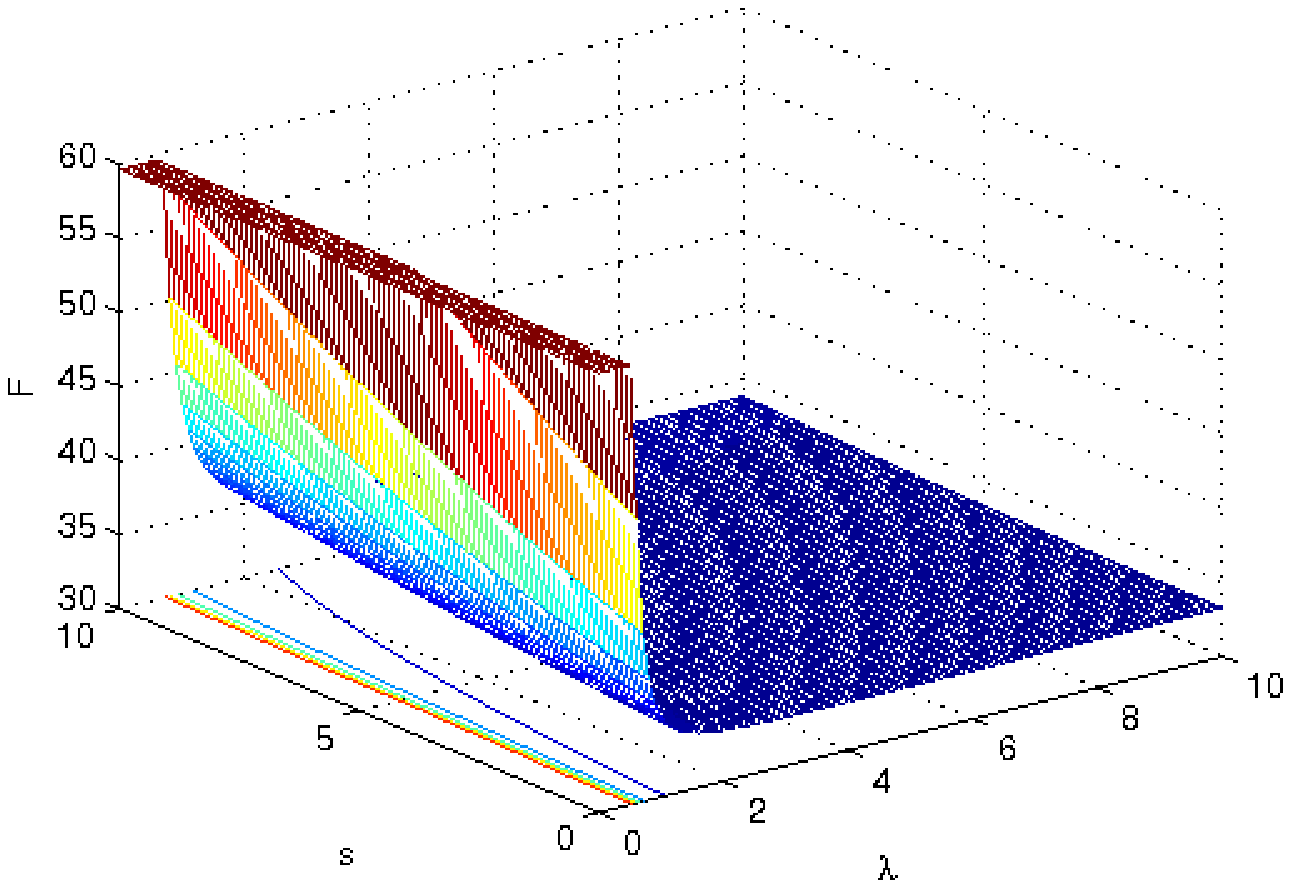}
\caption{Control with protected areas for the healthy prey, for fixed $\omega=5$.}
\label{fig:s_lambda}
\end{figure}

To make things clearer, consider introducing the protected areas for the healthy prey, i.e. let us give to $s$ nonzero values. In Figure \ref{fig:s_lambda}
we plot the population levels as functions of both $\lambda$ and $s$. Here the value of $\omega$ as said is kept at level $5$, and independently of
the fixed value of $s$ chosen,
we see that the equilibrium values of the infected, center frame, as a function of $\lambda$ has a similar behavior
as if no control were present, it raises up to a maximum and then decreases. As function of $s$ it is slightly decreasing.
Note that the maximum of infected with no disease control for $\omega=5$ and $\lambda=2$ is about $35$, Figure \ref{fig:no_contr} center.
When refuges for the healthy prey are
present the number of infected remains about the same for increasing values of $s$, Figure \ref{fig:s_lambda} center, probably indicating a
scarce effect of this measure to contain the disease propagation.

\begin{figure}[!htb]
\centering
\includegraphics[width=8cm]{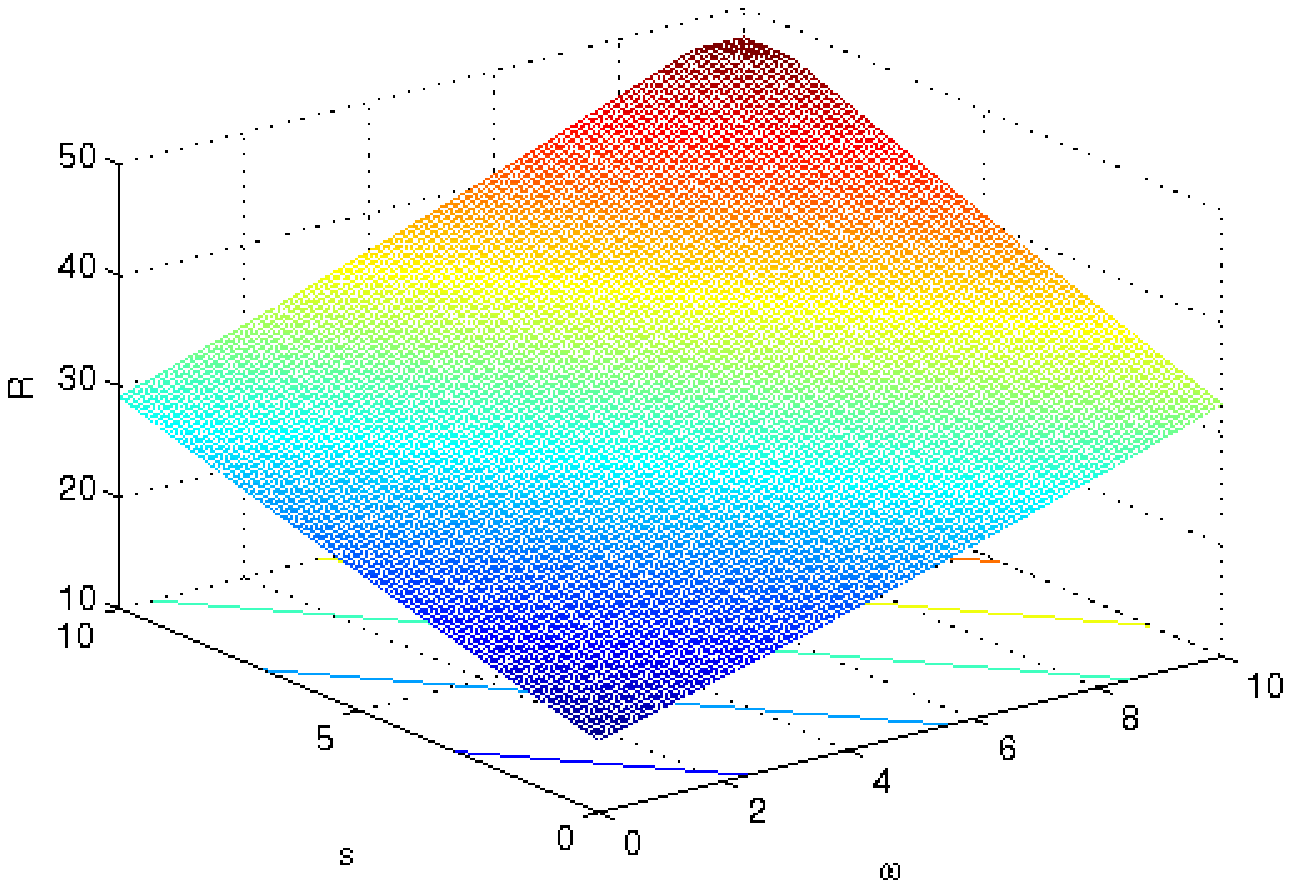}
\includegraphics[width=8cm]{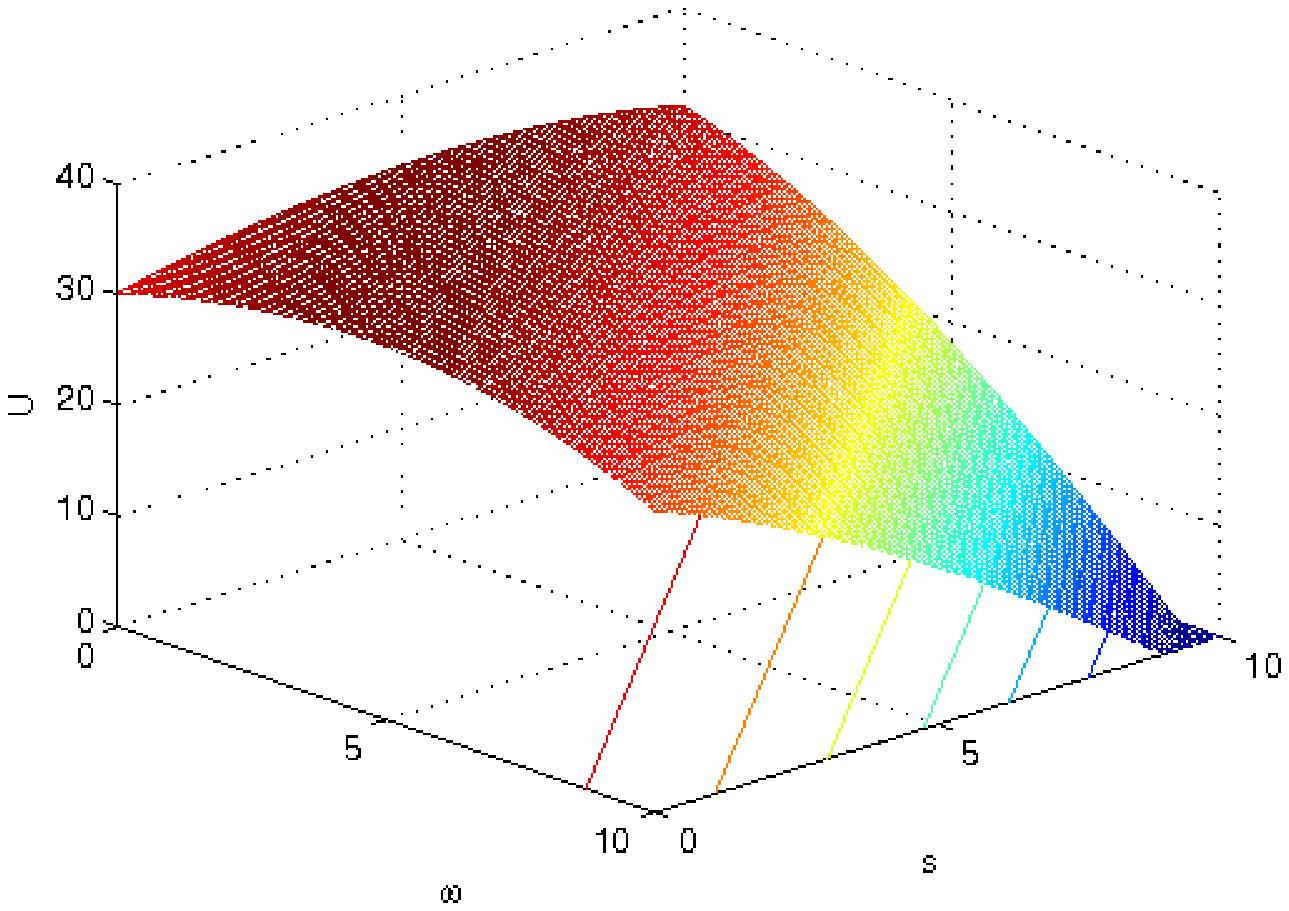}
\includegraphics[width=8cm]{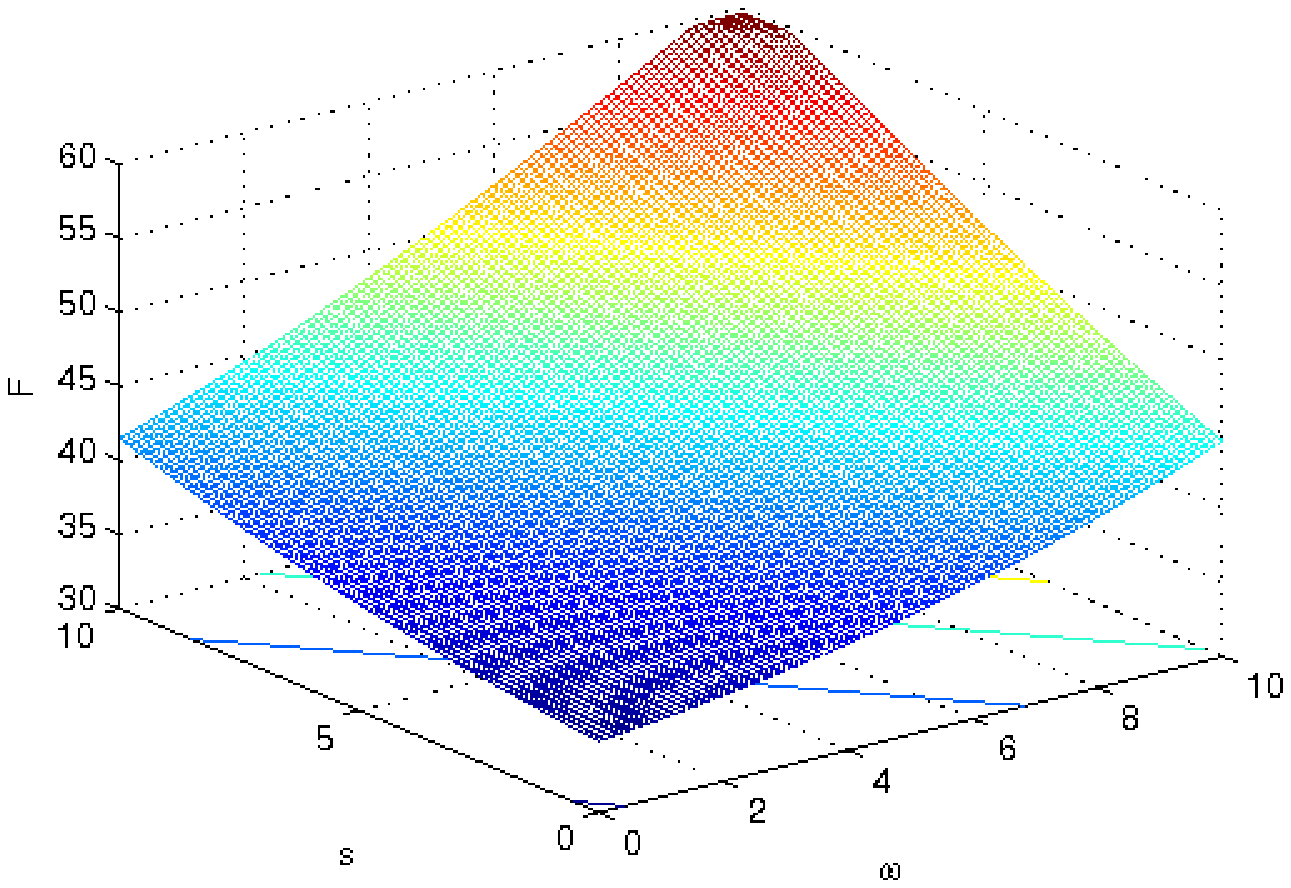}
\caption{Control with protected areas for the healthy prey, for fixed $\lambda=2$. For the infected the plot is shown under a different angle,
to show disease eradication for suitable values of the parameters.}
\label{fig:s_omega}
\end{figure}

If we study the same situation as a function of the parameter $\omega$, we have Figure \ref{fig:s_omega} center, for $\lambda=0.8$,
shown under a different angle to better indicate that for large values of the control $s$ and the recovery rate $\omega$ the disease gets
eradicated.
Now in Figure \ref{fig:s_omega} center we need to restrict the surface to the plane $\omega=5$. The resulting function decreases with
increasing $s$, starting from a value that for $s$ close to zero is comparable to the reference one.

If we compare the healthy prey equilibrium values, for $s \approx 0$ and $\omega=5$, Figure \ref{fig:s_lambda} top, the situation is similar to the case
of no disease control, Figure \ref{fig:no_contr} top. Supplying the healthy prey refuges, has the benefit that the equilibrium level of the latter
increase, e.g. for $s=10$ and $\lambda=10$ we find $R=10$, certainly higher than the level in Figure \ref{fig:no_contr} top for $\omega=5$
and for the corresponding value of $\lambda=10$.
Similarly we find that as a function of $\omega$ the equilibrium level surface is almost always above the level $R=10$, thus improving over
the case of no control, Figure \ref{fig:s_omega} top. In particular in this latter case the increase in number of healthy prey is quite dramatic.

\subsection{Comparison of the four different controls}
\begin{figure}[!htb]
\centering
\includegraphics[width=5.5cm]{U_s_lambda.eps}
\includegraphics[width=5.5cm]{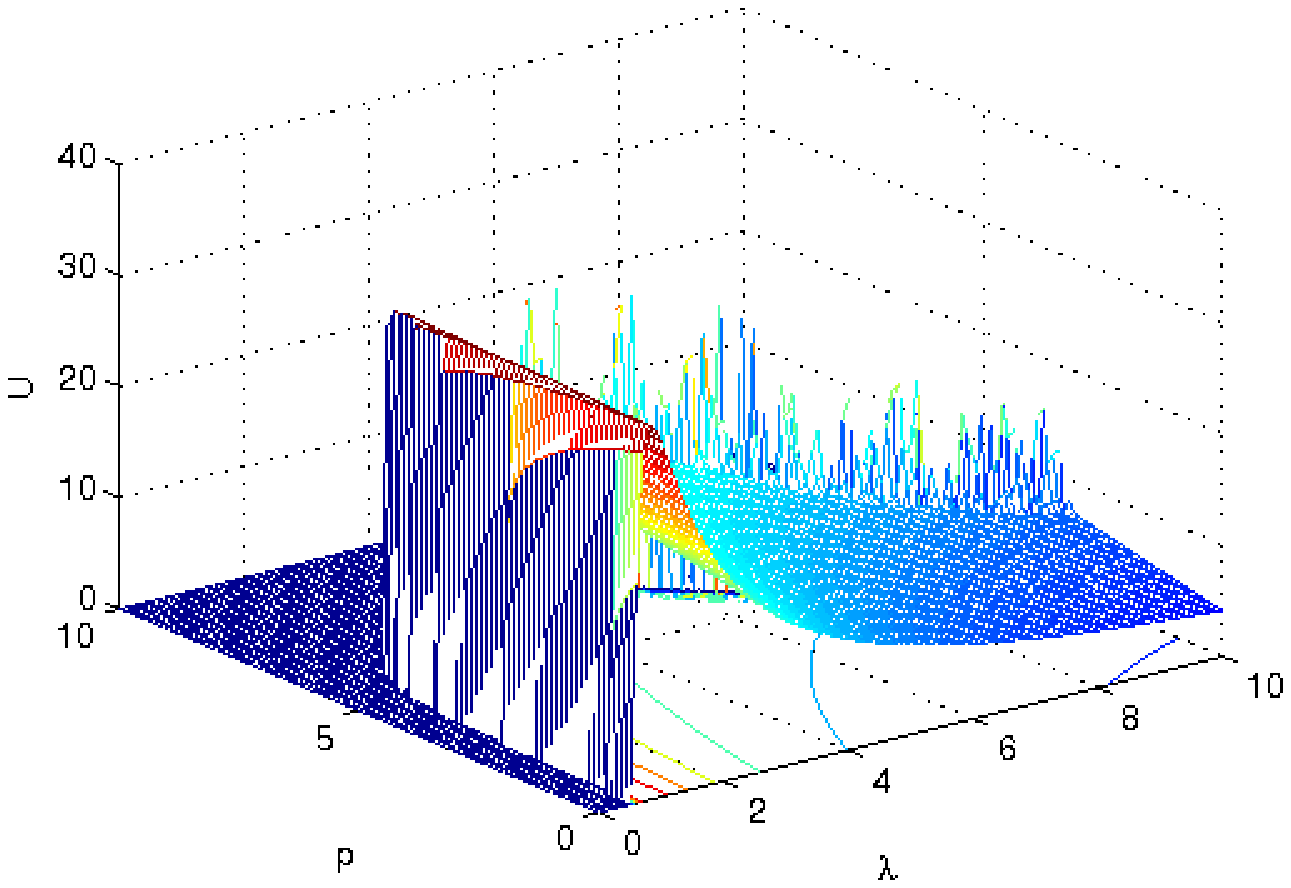}\\
\includegraphics[width=5.5cm]{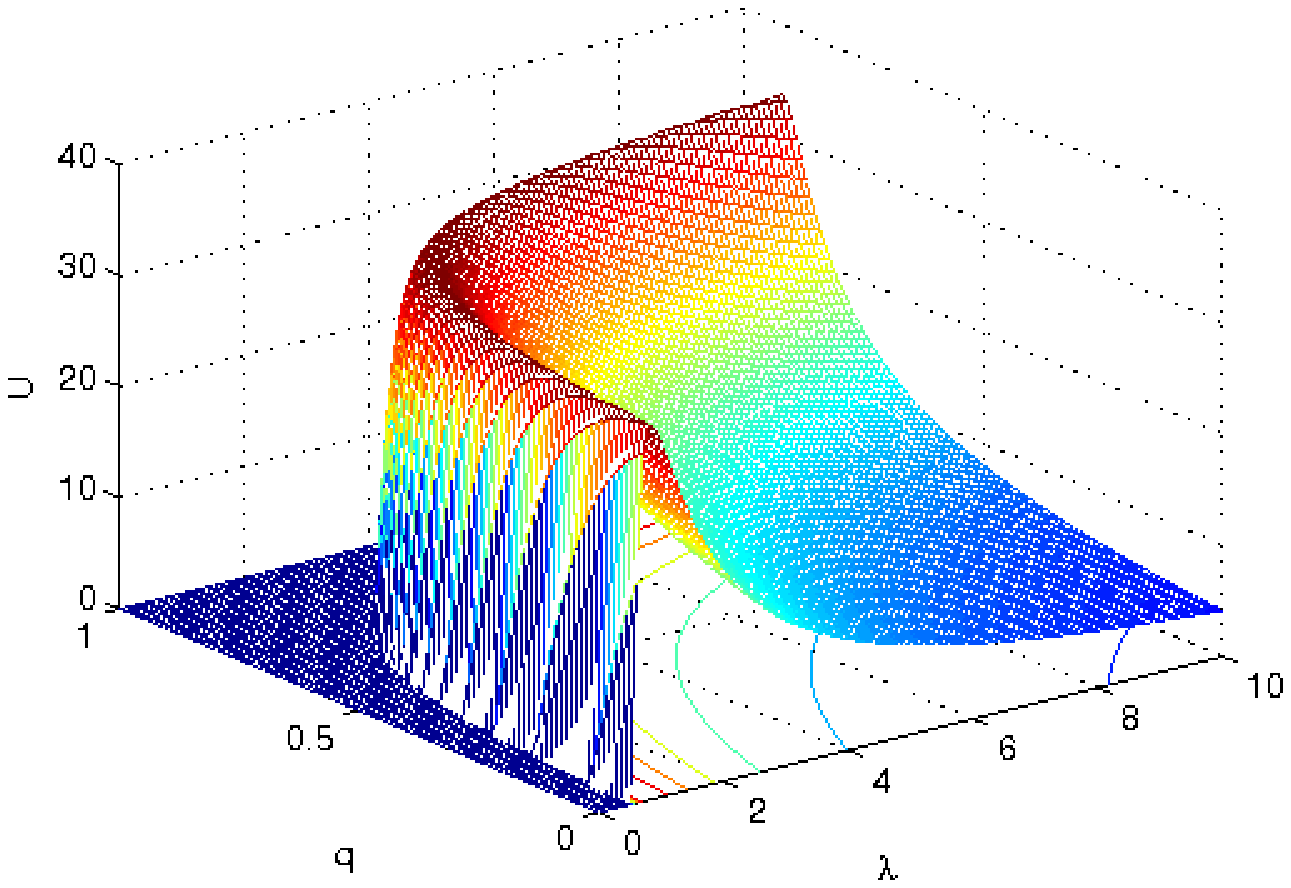}
\includegraphics[width=5.5cm]{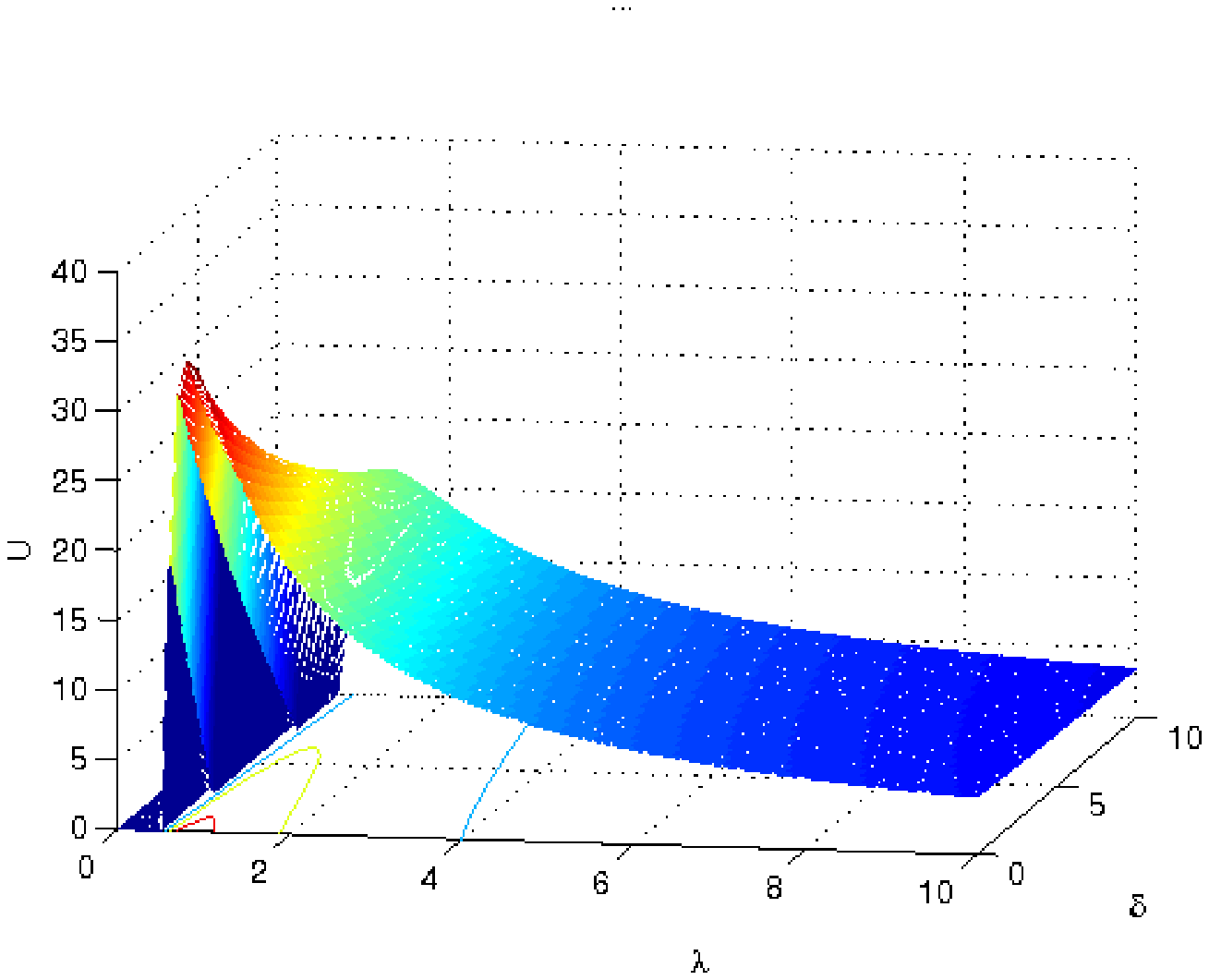}
\caption{Infectives as function of the various controls; left to right and top to bottom $s$, $p$, $q$ and $\delta$, for fixed $\omega=5$. Note that the spikes in the top right plot correspond to the situations in which the equilibrium
is ustable and the coexistence is attained through persistent oscillations.}
\label{fig:U_lambda}
\end{figure}

We now consider the four different controls. Plotting in the same frame the infected as function of
all of them and of $\lambda$, Figure \ref{fig:U_lambda}, left to right and top to bottom the controls
being $s$, $p$, $q$ and $\delta$, we observe some differences in the infected equilibrium levels.
As already discussed above, the parameter $s$ seems to lead in general to a rather higher prevalence, uniformly and independently
of the contact rate $\lambda$. A sufficiently high value of both $p$ and $q$, for not too high values of $\lambda$,
lead to disease eradication, e.g. $\lambda=1$ and $p=7$ or $q=0.7$.
However, both seem to have drawbacks: the ``inappropriate'' use of $p$ or $q$ leads to a high peak in
the prevalence, for $\lambda \approx 2$ and $q \approx 0.7$. This occurs throughout the possible ranges of the controls and of
the disease transmission rate, following the peaks in the two frames. The difference however is that the peak is rather steady when
the control $q$ is used, while it decreases slowly in case of $p$. So among these two controls, the refuge for the infected prey is
preferable. In fact, in this case
a choice of a large $p$ when $\lambda$ is also large leads to
persistent oscillations, as remarked earlier, Figure \ref{fig:oscill}, which correspond to the uneven
portion of the surface
in the upper right corner of Figure \ref{fig:U_lambda}, frame for the control $p$.

Culling markedly decreases the peak of the prevalence when $\lambda=0.8$, but it gives a much smaller range for which the disease is eradicated
compared to the use of $q$ and $p$. The ``zero level'' surface has a larger area indeed in the frames for the reduced contacts and
the refuges for infected prey controls than what we find in the frame for culling.
For large values of the transmission rate and low levels of the controls $p$, $q$ and $\delta$ the number of infected at equilibrium settles to about
the same value $U=10$. For larger implementations of these controls however, there is a marked difference. For $q$ the prevalence shoots up and only
for extreme values of the control it goes down and eventually disappears. When using culling, the infected equilibrium levels do not change much
even if high rates of abatement are employed. For the refuges for infected prey strategy, prevalence remains about the same, then there is a
regime of oscillatory behavior, and finally for larger values of the control the disease is eradicated.

\begin{figure}[!htb]
\centering
\includegraphics[width=5.5cm]{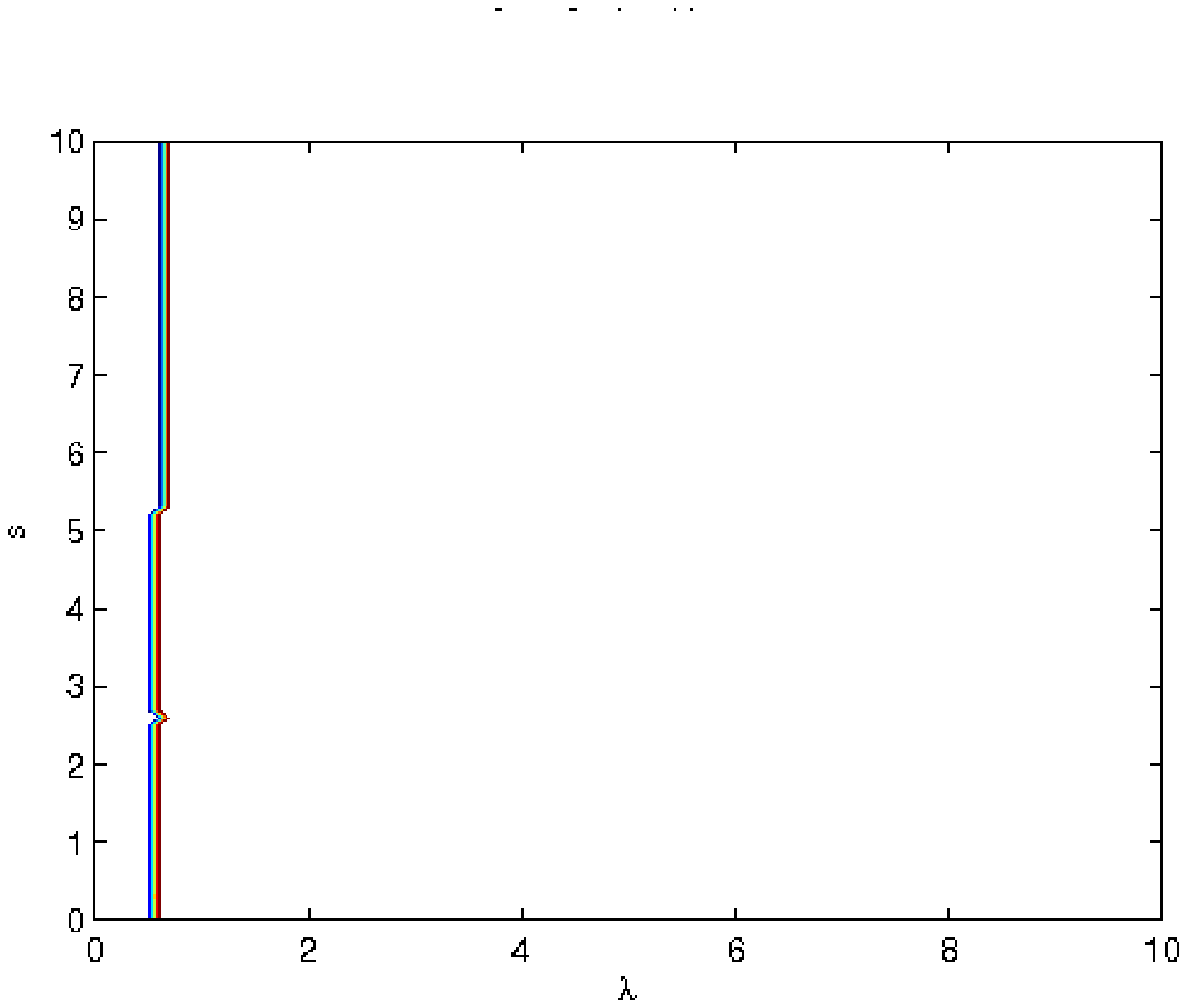}
\includegraphics[width=5.5cm]{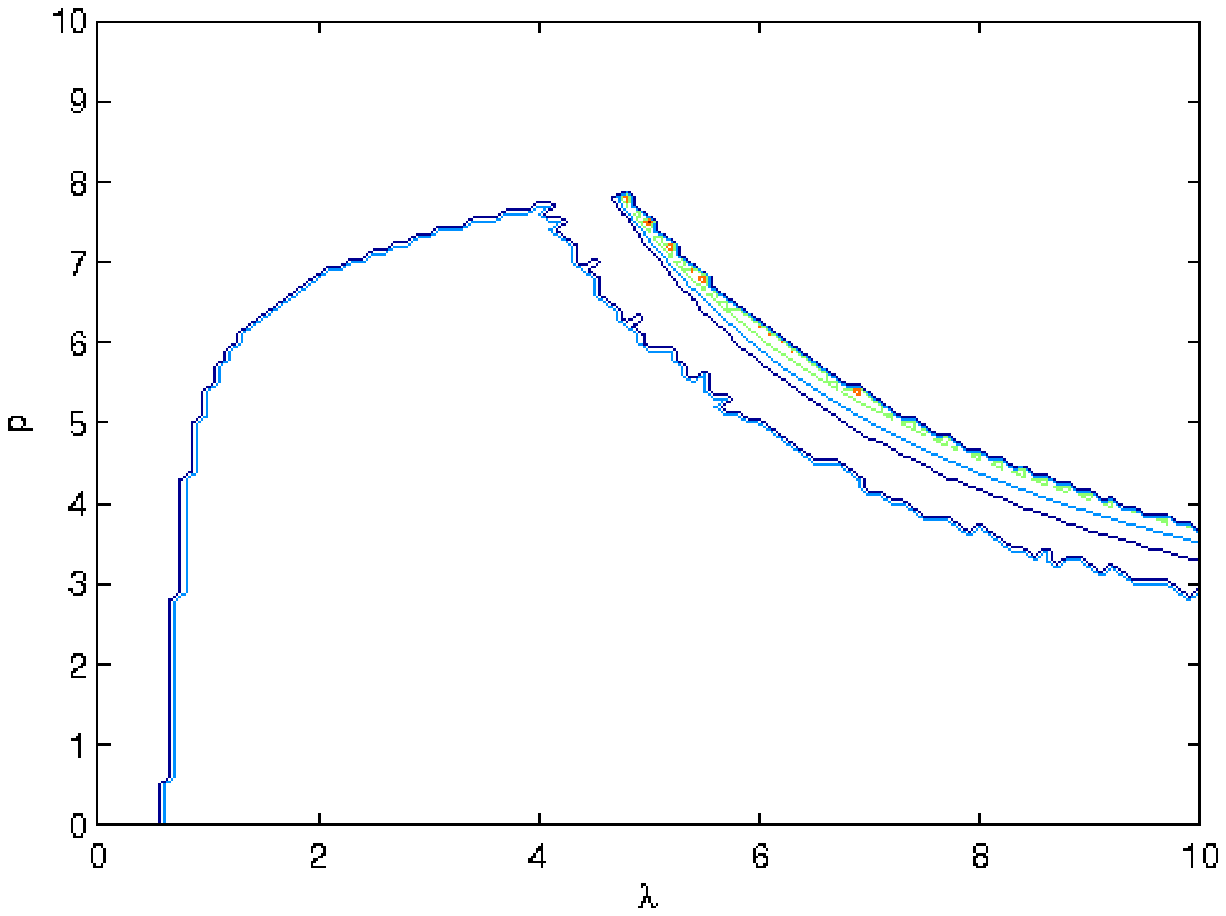}\\
\includegraphics[width=5.5cm]{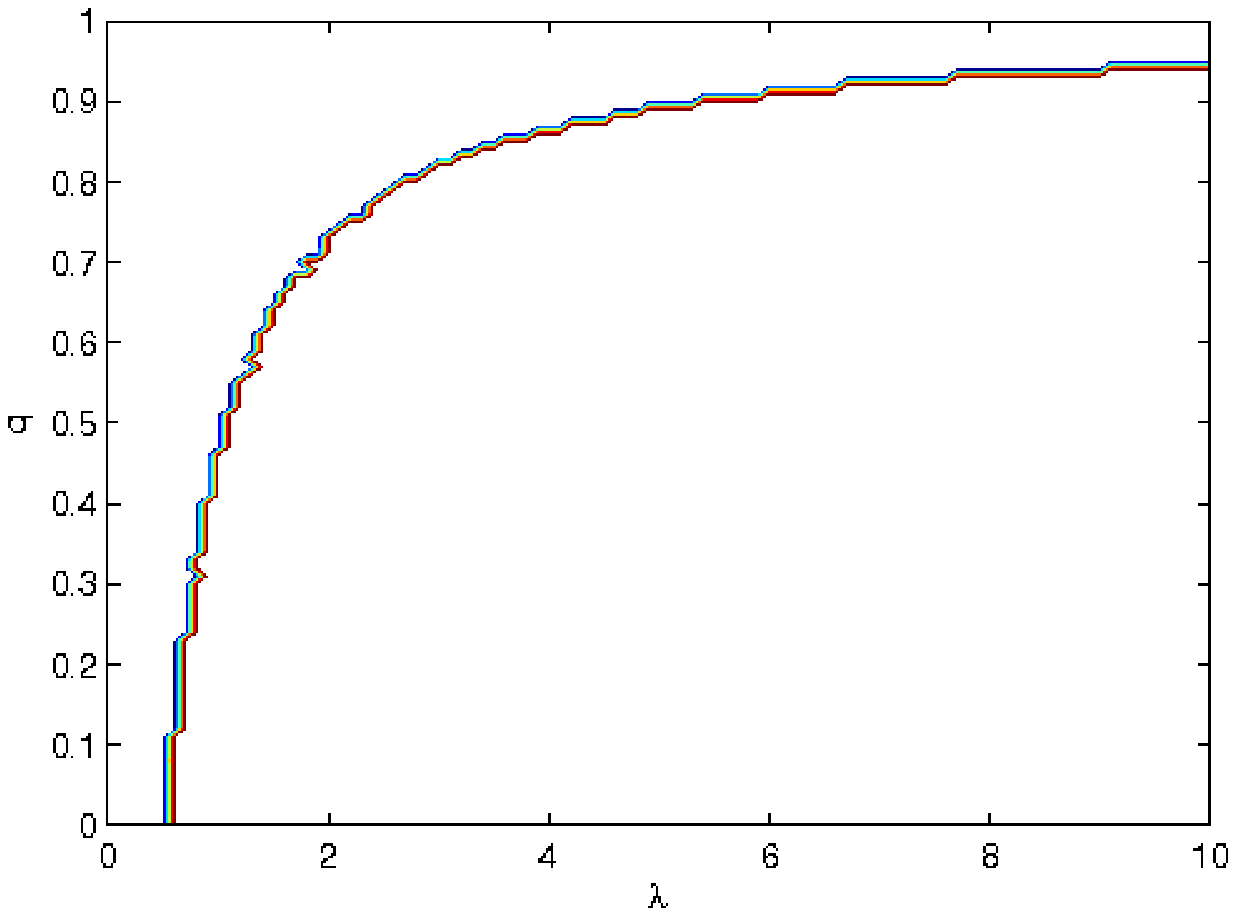}
\includegraphics[width=5.5cm]{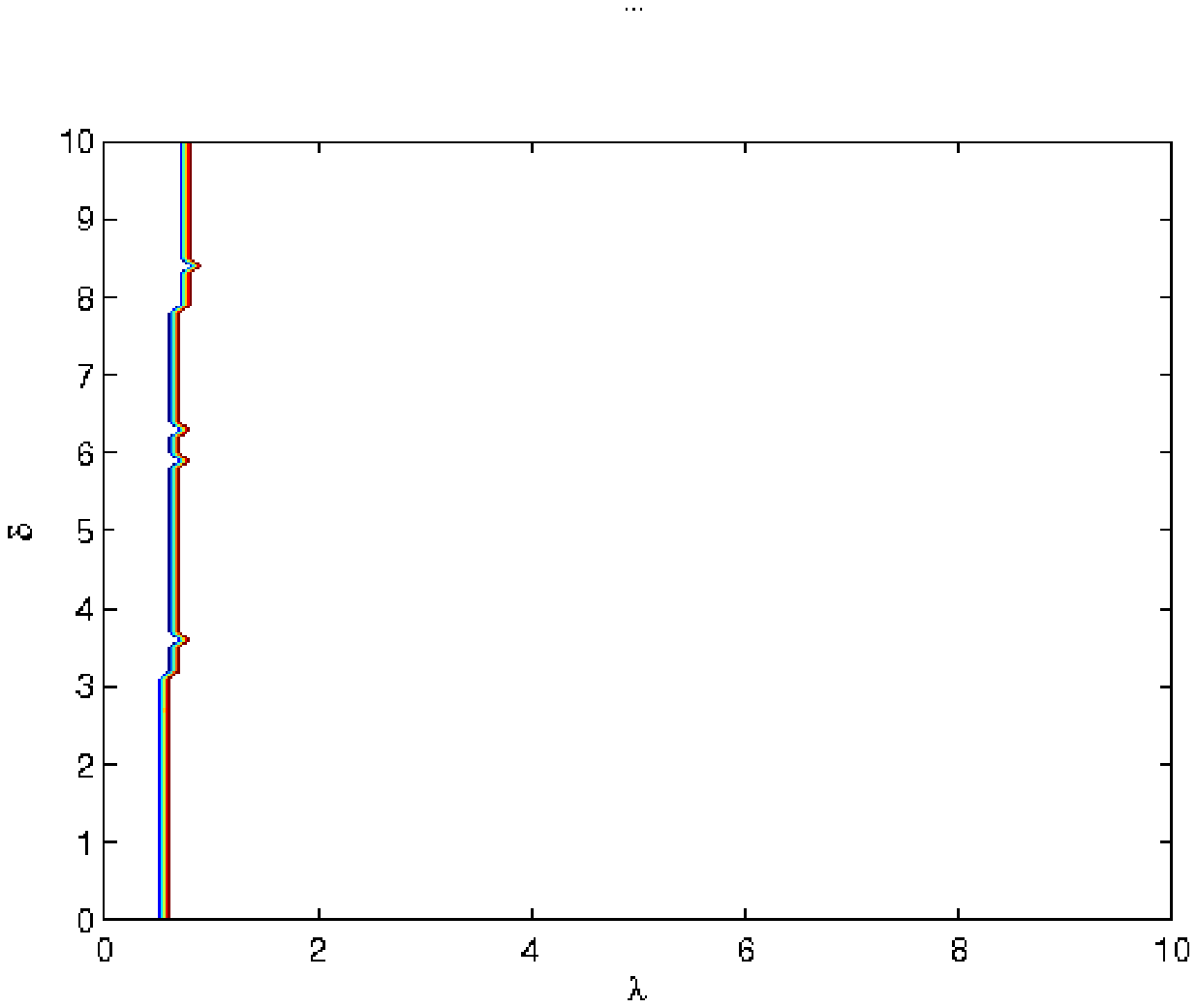}
\caption{Loci of the equilibria in the various controls-$\lambda$ parameter space.
Infectives levels are function of the various controls; left to right and top to bottom $s$, $p$, $q$ and $\delta$, for fixed $\omega=5$. In the top left and bottom right frames,
the region to the left of the vertical line is the disease-free equilibrium, to its right we have the endemic equilibrium.
Similarly in the other frames,
above the curve there is disease eradication, below the disease is endemic. In the plot with the control $p$ also the oscillatory
region is indicated, at the border of the previous two regions for high transmission rates.}
\label{fig:C_lambda}
\end{figure}

To better study the limit cycles, we plot in Figure \ref{fig:C_lambda} the parameter space of the controls used versus the disease transmission
rate. The curves in each plot separate the region in which the disease is eradicated, the one having as border the vertical axis, from the region
where the disease is endemic, the one bordering the horizontal axis. The region of the limit cycles appears only when the $p$ control is used,
at the interface of the two regimes, for large values of the contact rate $\lambda$.
The largest area for the disease-free equilibrium is therefore observed in case control is exercised through the parameter $p$.

\begin{figure}[!htb]
\centering
\includegraphics[width=5.5cm]{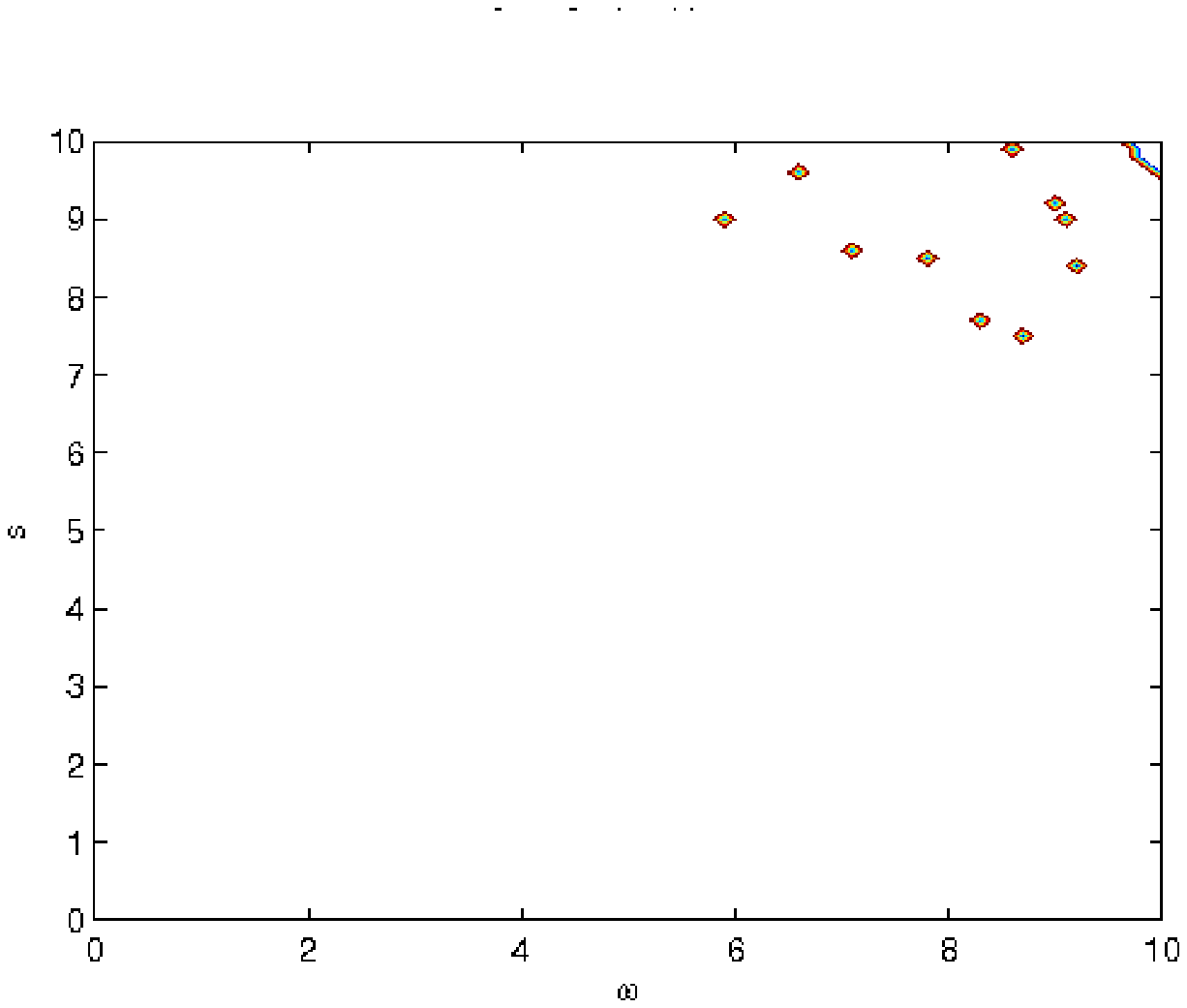}
\includegraphics[width=5.5cm]{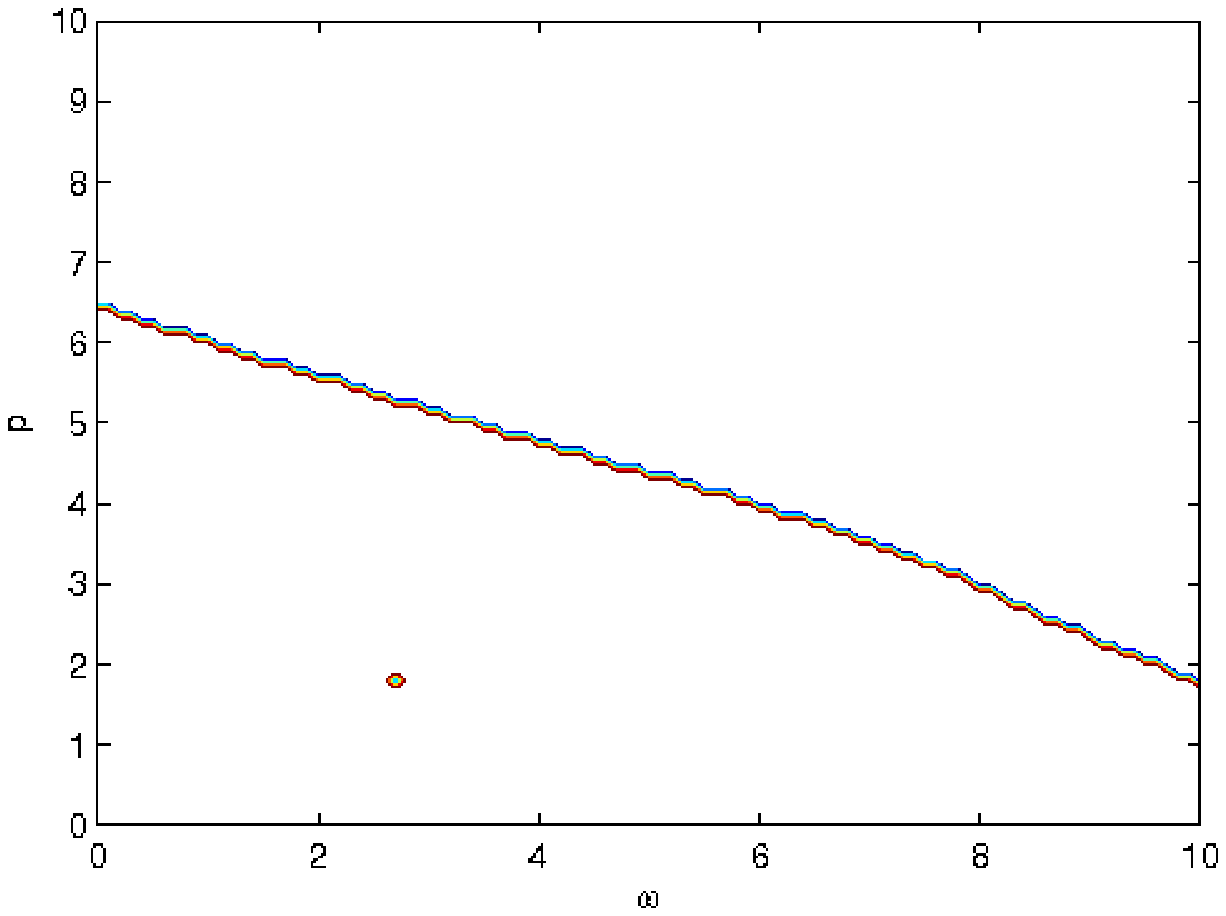}\\
\includegraphics[width=5.5cm]{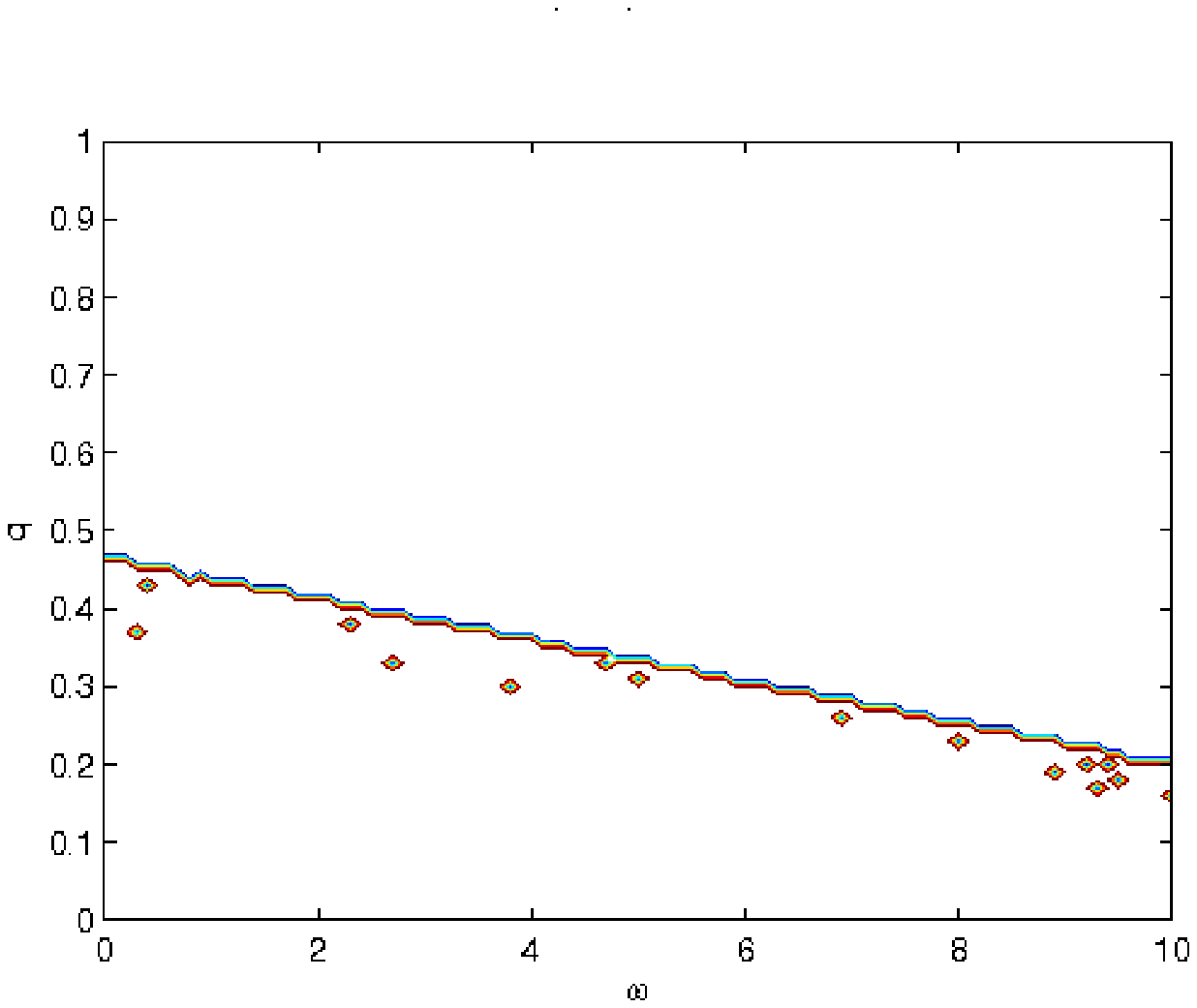}
\includegraphics[width=5.5cm]{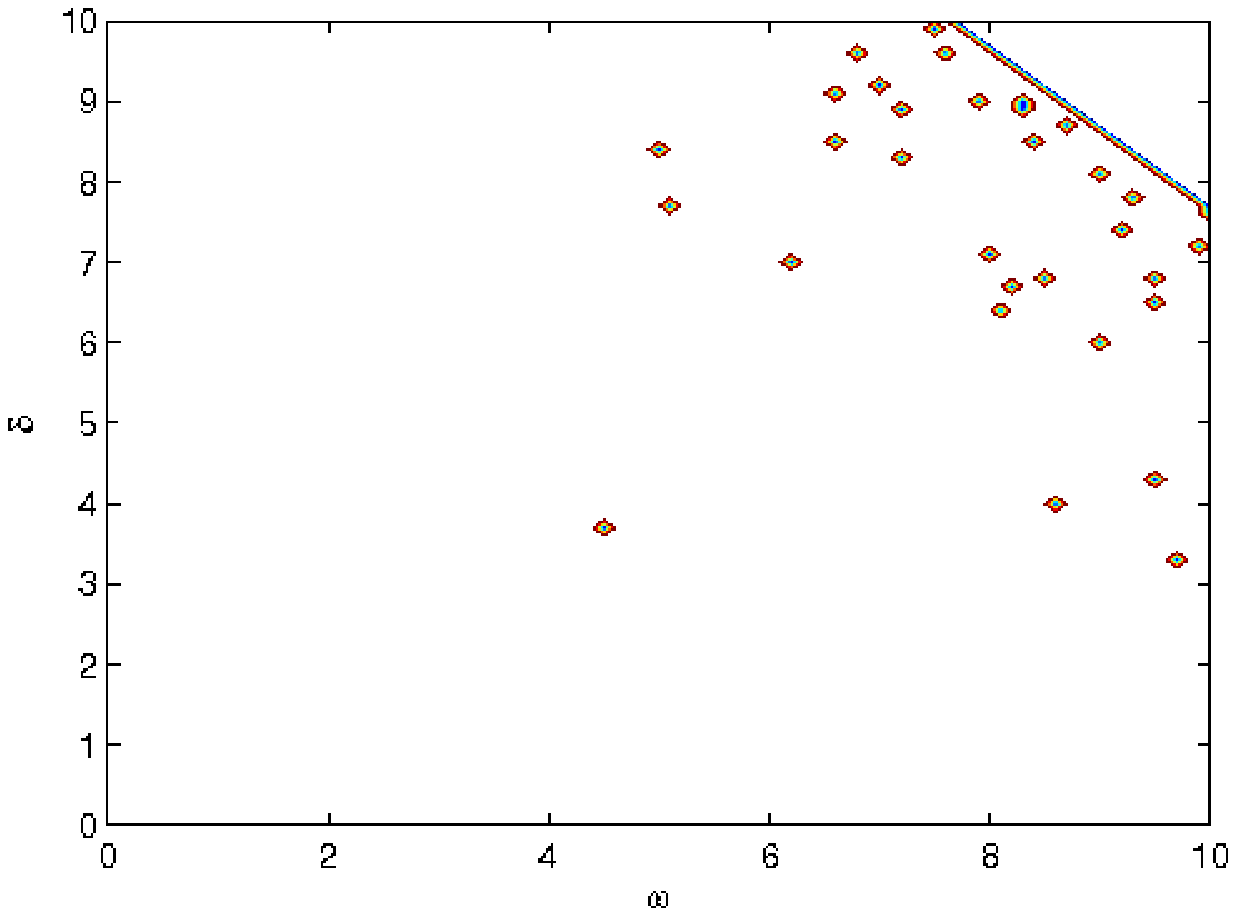}
\caption{Loci of the equilibria in the various controls-$\omega$ parameter space.
Infectives levels are function of the various controls; left to right and top to bottom $s$, $p$, $q$ and $\delta$, for fixed $\omega=5$.
The region containing the origin represents the endemic equilibrium.
For the $s$ control, the disease-free equilibrium region is a very small triangle in the top right corner.
The spots that occasionally appear correspond to very tiny oscillations, that can be disregarded.
The largest region in this parameter space providing disease eradication is given by the reduced contacts policy.}
\label{fig:C_omega}
\end{figure}

In Figure \ref{fig:C_omega}
we also compare the 
loci of the equilibria in the various controls versus the disease recovery rate parameter space.
Here the region containing the origin represents always the endemic equilibrium.
The $s$ control exhibits the smallest disease-free equilibrium region, the very small triangle in the top right corner.
Similarly to it behaves culling.
The reduced contacts and the refuge for infected prey controls have much larger regions where the disease is eradicated, with
the largest region apparently being provided by the former policy, recalling that $q$ is a fraction and cannot exceed 1.

\begin{figure}[!htb]
\centering
\includegraphics[width=5.5cm]{R_s_lambda.eps}
\includegraphics[width=5.5cm]{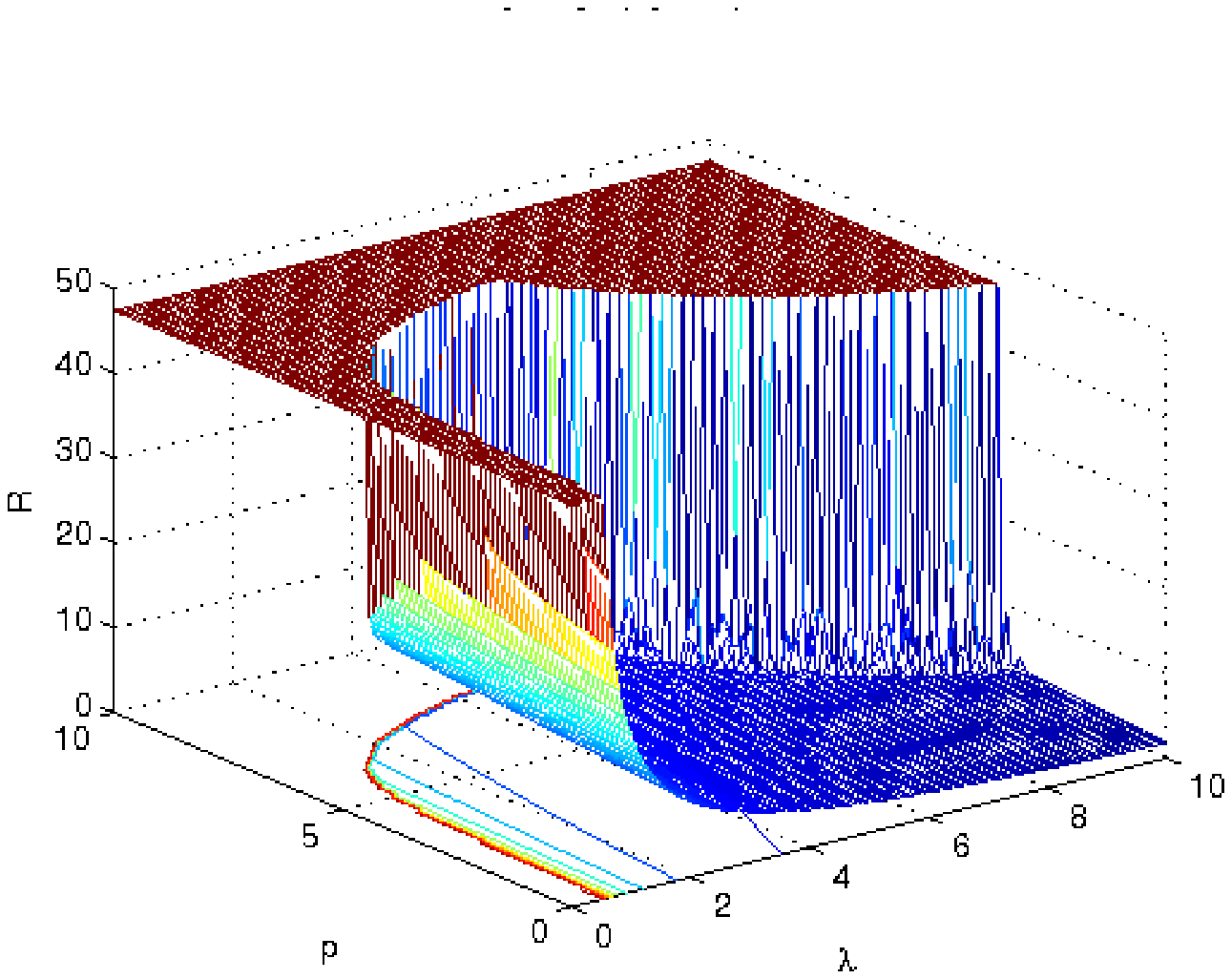}\\
\includegraphics[width=5.5cm]{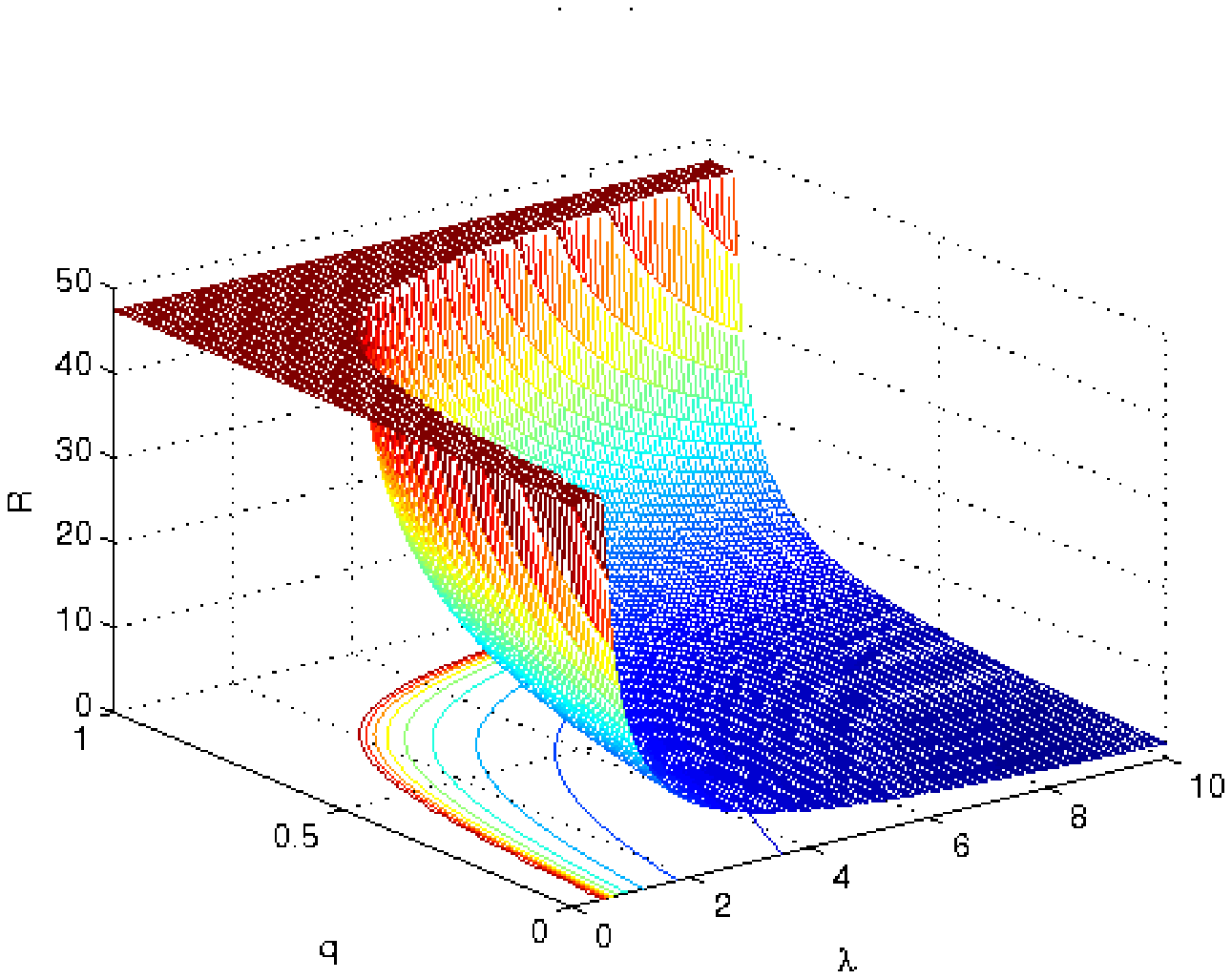}
\includegraphics[width=5.5cm]{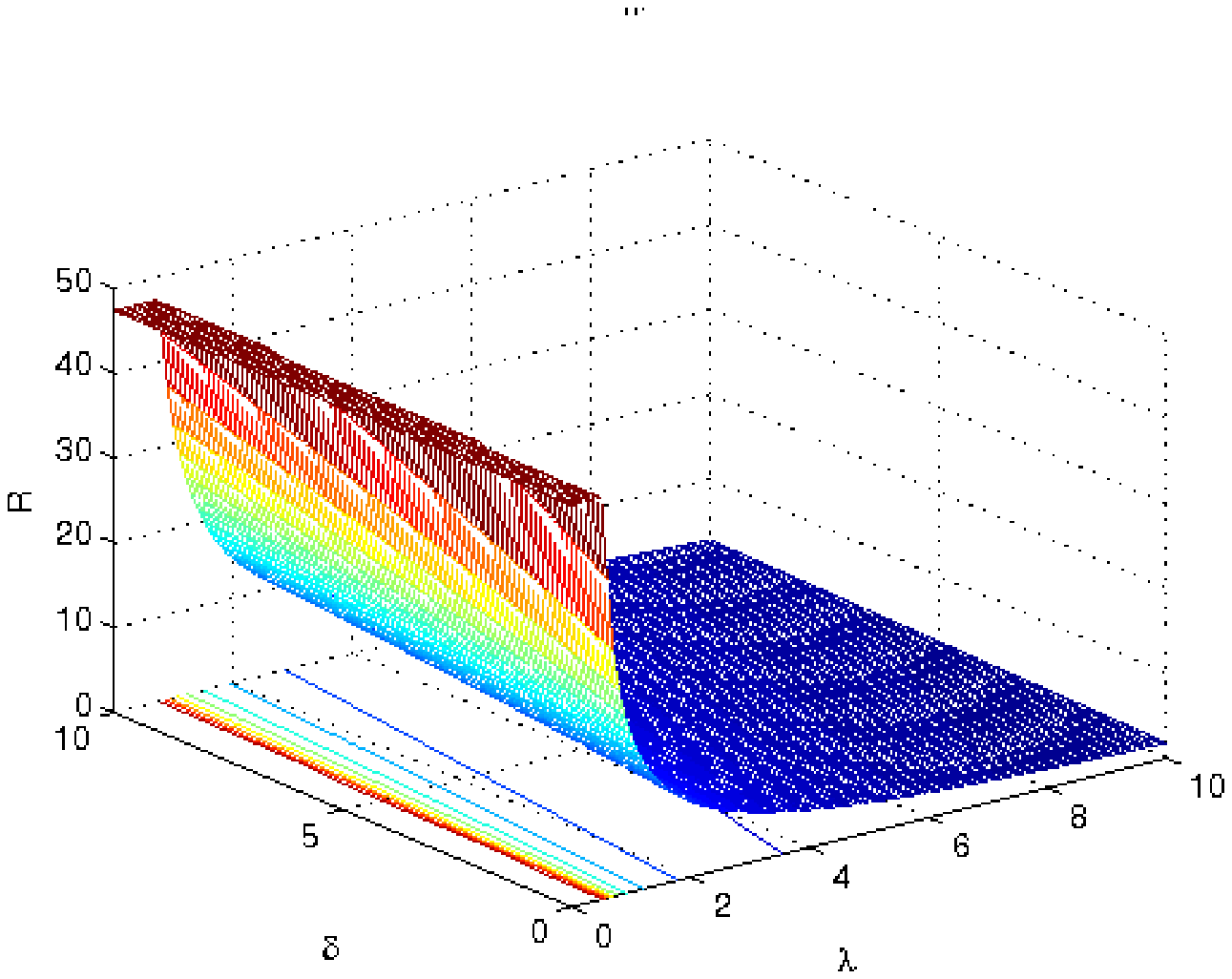}
\caption{Healthy prey as function of the various controls; left to right and top to bottom $s$, $p$, $q$ and $\delta$, for fixed $\omega=5$.}
\label{fig:R_lambda}
\end{figure}

\begin{figure}[!htb]
\centering
\includegraphics[width=5.5cm]{F_s_lambda.eps}
\includegraphics[width=5.5cm]{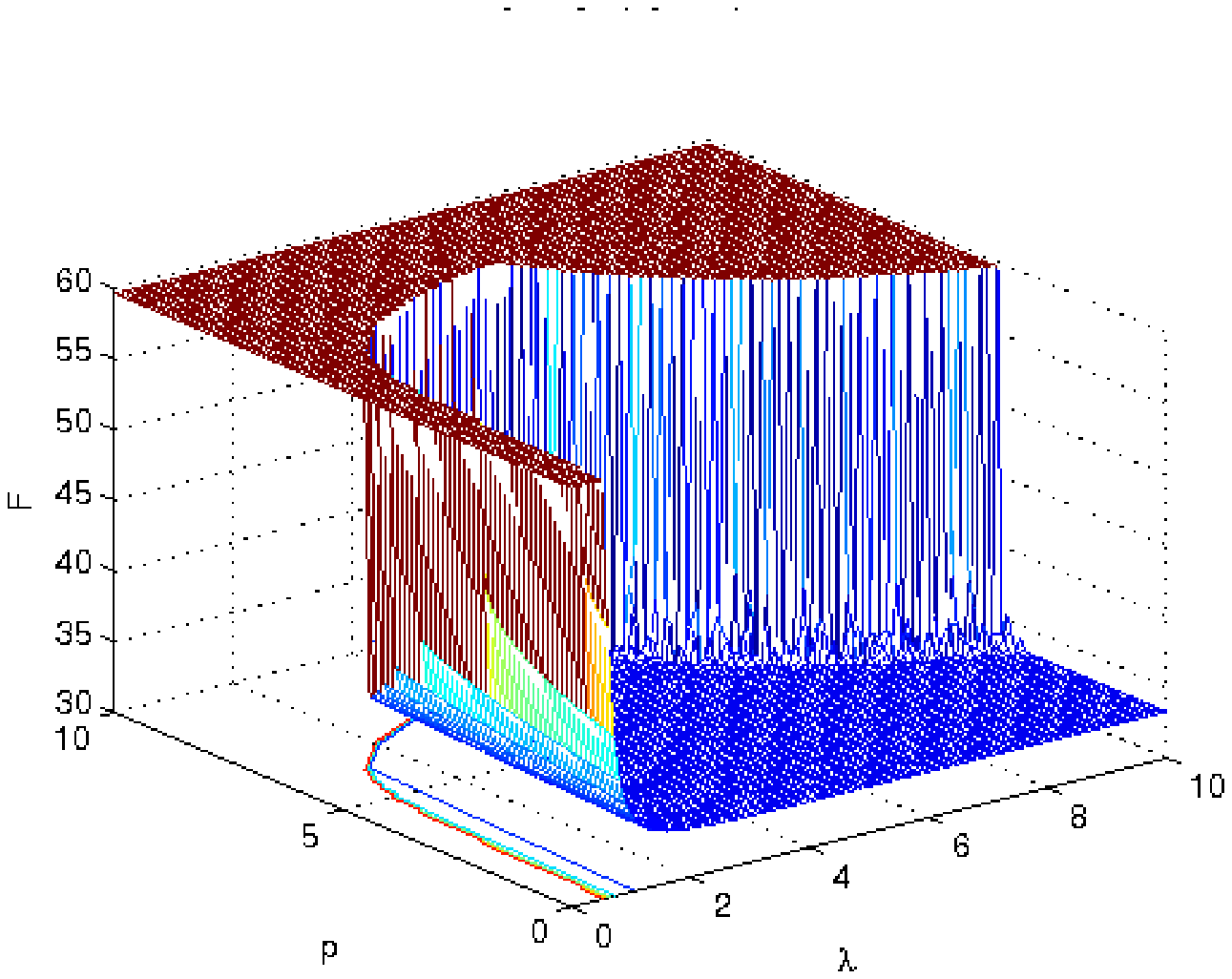}\\
\includegraphics[width=5.5cm]{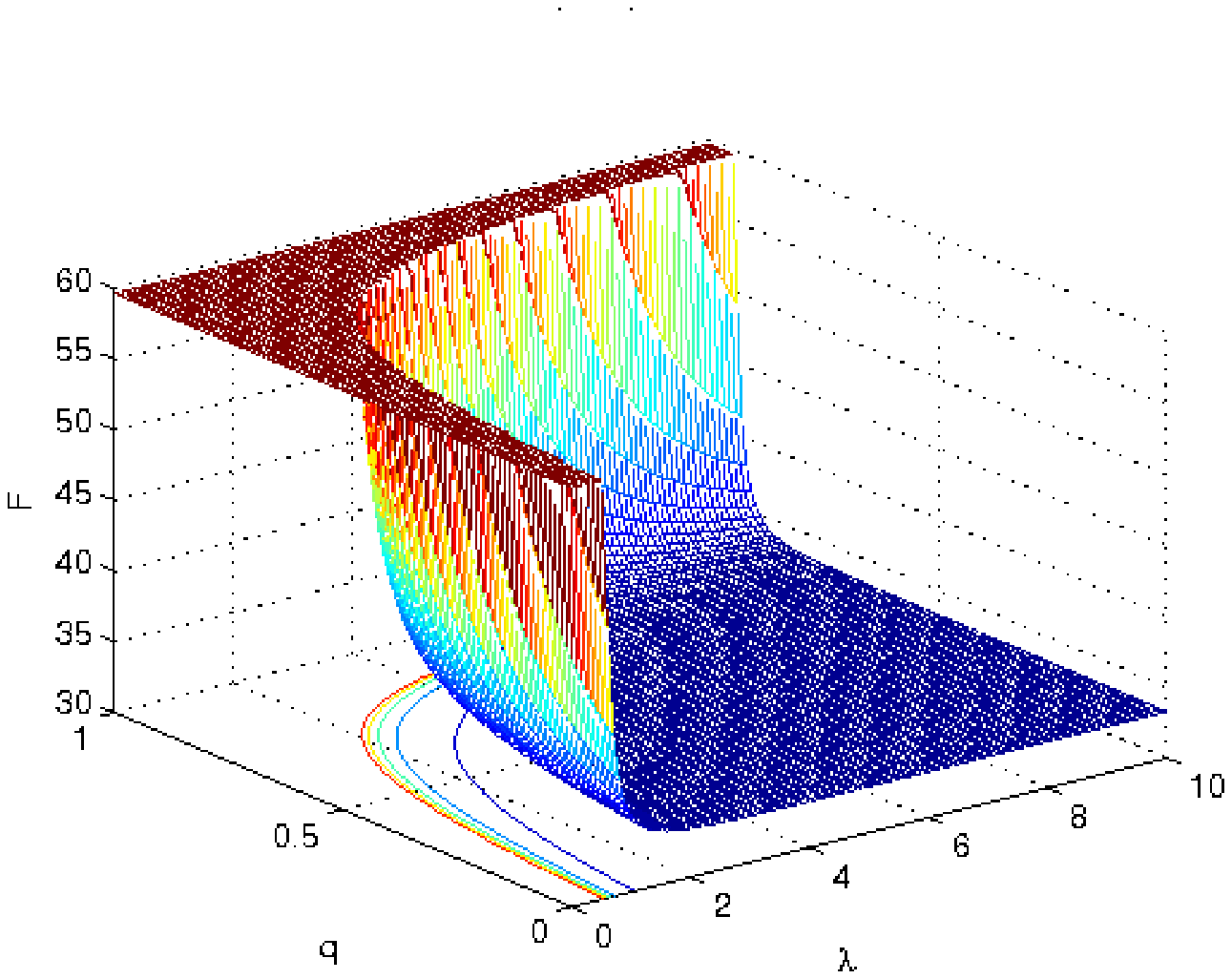}
\includegraphics[width=5.5cm]{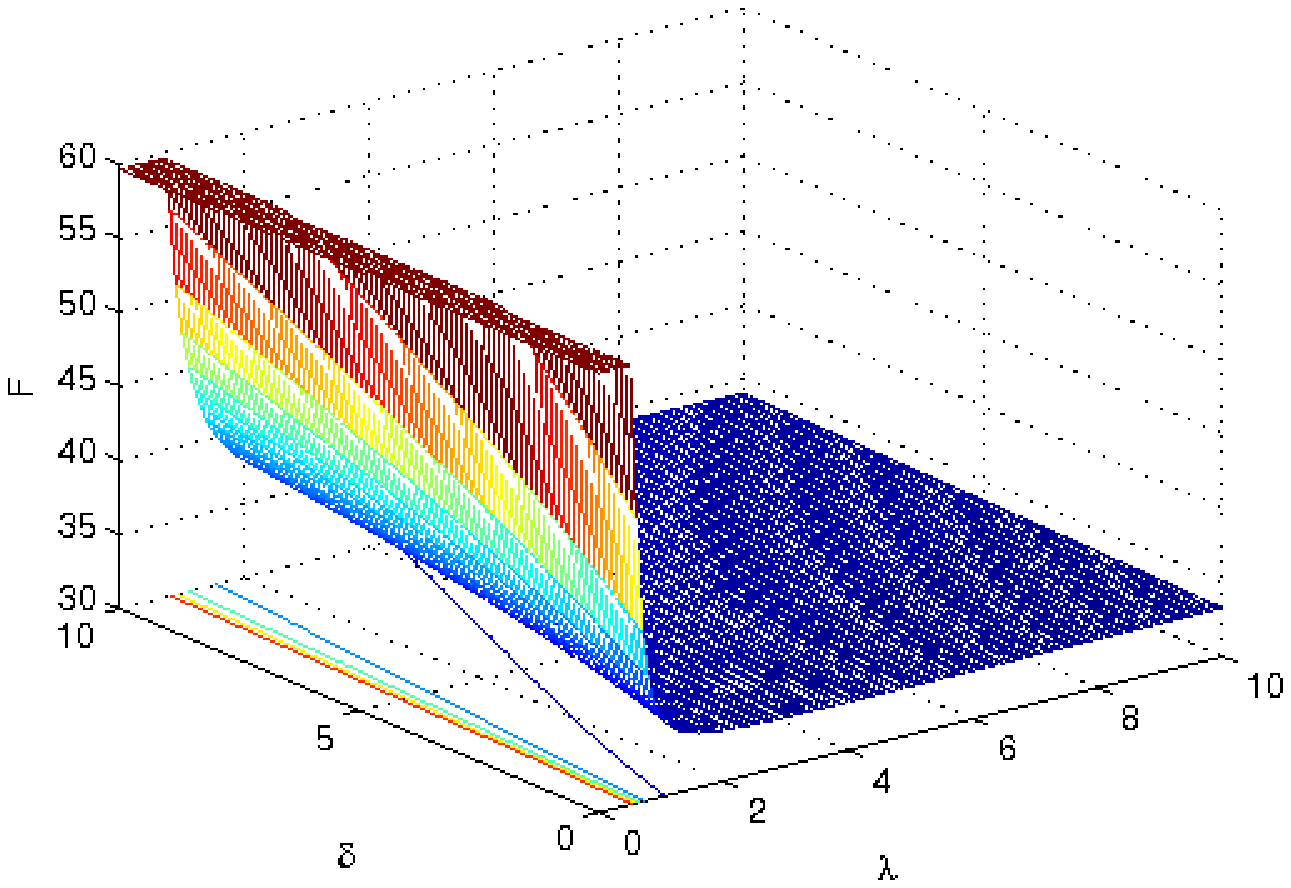}
\caption{Predators as function of the various controls; left to right and top to bottom $s$, $p$, $q$ and $\delta$, for fixed $\omega=5$.}
\label{fig:F_lambda}
\end{figure}

Comparing the healthy prey and predators levels, Figures \ref{fig:R_lambda} and \ref{fig:F_lambda}, similar conclusions can be drawn.
Culling and refuge
for the healthy prey seem to behave similarly to each other, certainly less effectively than the other two policies. Among these two,
as far as the healthy prey are concerned, it seems to be preferable not to use culling, since for large transmission rate $\lambda \approx 10$,
for high values of the control $s$, they have a value around 10,
while they attain much smaller values independently of the culling rate used.
The predators levels are instead about the same for both policies also for large $\lambda$.
The policies of refuges for the infected prey and of reducing the contact rate instead, when heavily implemented, i.e. for large values of
the parameters $p$ and $q$, boost both healthy prey and
predators populations levels, especially in presence of high transmission rates,
see the left top corners of the corresponding figures. A clear advantage is obtained by providing refuges for the infected prey, where they
are less able to transmit the disease, see the top right frames in both Figures \ref{fig:R_lambda} and \ref{fig:F_lambda}.

\subsection{Final considerations}
In summary it seems that no strategy is the best alone. A clear exception are the safety refuges for healthy prey, in that they do not seem to be
effective in controlling the disease levels and therefore should not be used. Selective culling on infected prey has adverse effects
on healthy prey and predators, but it is preferable to control through reduced contacts in terms of smaller disease prevalence. In presence
of a high transmission rate the best policy is to use refuges for the infected individuals, taking into account
however that an insufficient
use of this control may trigger persistent oscillations in the system.

Thus in this type of predator-prey ecoepidemic system with disease just in the prey,
for an endemic disease, the ecosystem with a place where some of the healthy individuals can be segregated
from coming in contact with disease carriers would exhibit the worst features to preserve the epidemics to spread.
Probably the most indicated strategies are providing areas for the infected prey where they cannot come in contact with the healthy ones,
Reducing the contact rate and culling seem instead to have mixed effects.
This result could possibly give some hints to field ecologists as how to fight diseases in wild populations, in case some artificial refuges
for the diseased individuals,
unreachable by the healthy animals, can be
provided in specific real-life situations.

\begin{acknowledgements}
EV is indebted to Prof. Cristobal Vargas for a useful discussion upon this matter, leading to the
analysis of the model with culling.
This research was partially supported by the project “Metodi numerici in teoria delle popolazioni”
of the Dipartimento di Matematica “Giuseppe Peano”.
\end{acknowledgements}



\end{document}